\documentclass[conference,compsoc]{IEEEtran}
\usepackage{multirow}
\usepackage{graphicx}
\usepackage{amsmath}
\usepackage[linesnumbered,ruled,vlined]{algorithm2e}
\usepackage{mathtools,amssymb,lipsum}
\usepackage{cuted}
\usepackage{makecell}
\usepackage{amsthm}
\usepackage{multicol}
\newtheorem{theorem}{Theorem}[section]
\newtheorem{lemma}[theorem]{Lemma}
\newtheorem{Claim}[theorem]{Claim}
\usepackage{wrapfig}
\usepackage{comment}

\ifCLASSOPTIONcompsoc
  \usepackage[nocompress]{cite}
\else
  \usepackage{cite}
\fi
\ifCLASSINFOpdf
\else
\fi
\begin{document}
%
\title{Secure Computation over Encrypted Databases}


\author{\IEEEauthorblockN{Tikaram Sanyashi}
\IEEEauthorblockA{Computer Science and Engineering\\
IIT Bombay,Mumbai,India\\
Email: tikaram@cse.iitb.ac.in}
\and
\IEEEauthorblockN{Bernard Menezes}
\IEEEauthorblockA{Computer Science and Engineering\\
IIT Bombay,Mumbai,India\\
Email: bernard@iitb.ac.in}}



%


\maketitle


\begin{abstract}
Sensitive applications running on the cloud often require data to be stored in an encrypted domain. To run data mining algorithms on such data, partially homomorphic encryption schemes (allowing certain operations in the ciphertext domain) have been devised. One such line of work yields schemes for secure \textit{k-nearest neighbors} computation that is designed to provide both \textit{Data Privacy} and \textit{Query Privacy}. Enhancements in this area further ensure that the data owner approves each query issued by a query user before the cloud server processes it. In this work, we describe an attack that invalidates the \textit{key confidentiality} claim, which further invalidates the \textit{Data Privacy} claim for these schemes. We show that a query user can specially tailor a query to extract information about the secret key used to encrypt the data points. Furthermore, the recovered secret information can be used to derive all the plaintext data points breaking \textit{data privacy}. We then suggest enhanced encryption schemes that make such attacks on \textit{data privacy} impossible while incurring meager additional costs in performance. 
\end{abstract}


%
\IEEEpeerreviewmaketitle

\section{Introduction}
Cloud Service Providers (CSPs) empower end-users with several desirable facilities allowing quick deployment with low costs, greater performance, reliability, scalability, high storage capacity, and much more. Enterprises like Google, Microsoft, Amazon, IBM, etc., actively rent out computing resources as a service.
This practice gives the CSPs physical control over the data stored, and the end-user needs to trust the CSP not to misuse or leak it. Solving the security and privacy issues that cloud computing is riddled with has therefore become a topic for extensive research.

To address the data privacy concerns, end-users may choose to encrypt their data before storing it on the cloud. Many standard and secure encryption techniques like the AES \cite{DaemenRijmen} exist to help achieve this. However, this reduces the cloud to a mere storage repository, precluding data owners from using powerful computation capability of cloud servers to query the data for extracting meaningful knowledge and patterns. 

One possible solution is the use of homomorphic encryption that allows computation in the encrypted domain. Although practically realizable Fully Homomorphic Encryption schemes (that allow generalized computation over the encrypted domain) are yet to be devised, there exist computationally efficient schemes that allow only certain homomorphic operations and hence, secure and fast computation for certain functions. These are tailored to certain applications, such as secure computation of the \textit{$k$-nearest neighbors} ($sk$-NN) of a given query point from an encrypted database.

Numerous works \cite{Choi_Sunoh_Gabriel_Ghinita,Yao_Bin_Feifei_Li,WongCheung},
have focused on finding a solution to this problem. Earlier ones include that of Liu et al. \cite{ChenLiu} that proposes a method using orthogonal matrices to calculate $sk$-NN. This was later improved by Wong et al. \cite{WongCheung} through the use of Asymmetric Scalar Product-preserving Encryption (ASPE) to achieve better security. 
Later, other works \cite{Xu_Huiqi_Shumin_Guo,Yiu_Man_Lung_Ira_Assent} have also presented ways of performing $sk$-NN calculation on the cloud service provider (CSP). However, all of these stipulate that a data owner (DO), one with the ownership of the encrypted data in the cloud, and query users (QU), the clients that perform queries on that data, are mutually trusted entities - an assumption that is not very realistic. 

More recently, Zhu et al. have presented a new architecture for $sk$-NN computation in \cite{ZhuHuang} considering each party as semi-honest -- meaning each entity strictly follows the protocol and returns the correct computation results, but at the same time, they try to infer as much information of other parties as possible based on the data they receive and hold. They consider the DO and QU as mutually exclusive sets, and the QU is not fully trusted. Secret information is therefore not shared with the QU. The encryption technique given in \cite{ZhuHuang} claims to provide the following four properties: 

(A) \textbf{Data Privacy against CSP}: Privacy of the DO's data should not be compromised to the CSP.

(B) \textbf{Key Confidentiality against QU}: DO's key remains confidential from the CSP and QU's. QU's should not get any information about the secret key even if QUs collude. 

(C) \textbf{Query Privacy against DO and CSP}: QU's query remains confidential from DO and CSP throughout the $k$-NN computation.

(D) \textbf{Query Controllability}: Without the DO's approval, QU should not be able to make any new queries.

However, the \textit{Query Controllability} claim of the scheme has been proven invalid by Singh et al. in \cite{SinghKaul}. 
\cite{SinghKaul} also provides an improved encryption scheme with the added property of  \textit{Query Check Verification}. 

Our contributions in this paper are as follows:

\begin{itemize}
    \item We investigate the schemes in \cite{ZhuHuang} and \cite{SinghKaul}, with focus on their \textit{key confidentiality} and \textit{data privacy}.
    \item We demonstrate a Level 1 Attack scenario that breaches the \textit{key confidentiality} claim in \cite{ZhuHuang}. We show that a QU can tailor its query to extract information about the secret key used to encrypt the data points. 
    \item We also describe an enhanced attack in the Level 2 Attack scenario where the querying user using the secret information derived in the Level 1 attack can extract all the plaintext points, thus violating the data privacy claim of \cite{ZhuHuang}. It can be further extended to invalidate the query privacy claim of the encryption scheme.
    \item A similar attack can be launched to breach the \textit{key confidentiality}, \textit{data privacy}, and \textit{query privacy} feature of the scheme in \cite{SinghKaul}. However, this attack can be launched only if the attacker resides inside the cloud -- which is an absolutely valid assumption in the cloud computation scenario.
    \item We suggest a new scheme with a similar underlying structure that makes breaching \textit{key confidentiality} and thus breaking \textit{data privacy} and \textit{query privacy} attack impossible while incurring very low additional performance costs.

\end{itemize}

The rest of the paper is organized as follows: Section \ref{Back} contains the background material, including an outline of the scheme in \cite{ZhuHuang} and \cite{SinghKaul}. Section \ref{attack_mod} demonstrates the attack on \cite{ZhuHuang} that renders its \textit{key confidentiality} and \textit{data privacy} claim invalid; this attack can also be extended to attack the scheme in \cite{SinghKaul}. Section \ref{Proposed_scheme} contains proposed enhanced encryption scheme for $sk$-NN computation preserving data privacy of the encryption scheme. 
Section \ref{results} compares the performance of the proposed encryption with that of the previous encryption schemes. Section \ref{Rel_work} contains related work and finally, Section \ref{conclusion} contains the concluding remarks.

Table \ref{Notation} contains the key mathematical notation used in the paper.

\begin{table}[h!] 
\centering
\caption{Table of Notations}
\begin{tabular}{|c|l|}
\hline
\hspace{0.3em} Notation \hspace{0.3em} & Meaning \\
\hline
\textbf{A} - \textbf{Z}  & matrices \\ 
\textbf{a} - \textbf{z}  & points  \\ 
a - z                    & constants \\ 
\hline
$\boldsymbol{\mathcal{D}}$  & set of data points \\ 
$\textbf{p}_i$          & $i^{th}$ data point \\
$\textbf{q}$            & query point \\ 
$d$            & dimension of data and query point \\
\hline
$\boldsymbol{\mathcal{D}}'$  & encrypted data set $\boldsymbol{\mathcal{D}}$ \\ 
$\textbf{p}_i'$          & encrypted data point $\textbf{p}_i$ \\
$\textbf{q}'$            & encrypted query point $\textbf{q}$ \\ 
$\mathbf{M}$    & secret invertible matrix \\
$r$             & random number $>0$ for query encryption\\
\hline
$\Vert \boldsymbol{p}_i\Vert$& $\sqrt{ \sum_{l=0}^{n} p_{il}^2}$ \\
$ED(\boldsymbol{p}_i,\boldsymbol{q})$     & $\sqrt{\sum_{l=0}^{n} (p_{il}-q_l)^2}$ \\
\hline
$c, \epsilon$             & security parameters $> 0$\\
$\textbf{s}$             & $(d+1)$-dimensional point of reals \\
$\boldsymbol{\tau}$          & $c$-dimensional point of reals \\
$\pi$           & permutation function on $\eta$ positions \\
$\textbf{v}$             & $\epsilon$-dimensional point of reals \\
$pk$            & Paillier encryption public key \\
$sk$            & Paillier encryption private key \\
$\boldsymbol{a}^{(q)}$ & encrypted query point from which $\textbf{q}'$ can be derived\\
$\beta_1,\beta_2$ & random query encryption constants\\
$\textbf{r}^{enc}$ & $c$-dimensional encrypted point\\
$\textbf{r}^{dec}$ & $c$-dimensional decrypted point\\
$\boldsymbol{\sigma}$ & $d$-dimensional point with real elements \\
$\boldsymbol{w}$ & $(c+\epsilon)$-dimensional weight point \\
$\boldsymbol{b}$ & $(c+\epsilon)$-dimensional bit string  with initial $c$ bits as $1$'s \\
& and remaining $\epsilon$ bits with at least one $0$ and one $1$ bit.\\

\hline


\end{tabular}
\label{Notation}
\end{table}

\section{Background}
\label{Back}
We denote the database of $d$-dimensional tuples in plaintext by $\boldsymbol{\mathcal{D}} = (\boldsymbol{p}_1,  \boldsymbol{p}_2, \cdots, \boldsymbol{p}_m)$ where the $i^{th}$ tuple is represented by its components as $(p_{i_1},  p_{i_2}, \cdots ,p_{i_d})$.  This is encrypted and stored in the cloud as $\boldsymbol{\mathcal{D'}} = (\boldsymbol{p}_1', \boldsymbol{p}_2', \cdots  ,\boldsymbol{p}_m')$ . The $i^{th}$ encrypted tuple is increased in dimension by ($1+c+\epsilon$) denoted as $(p_{i_1}' , p_{i_2}' , \cdots , p_{i_(d+1+c+\epsilon)}' )$ . Here, positive integers $c$ and $\epsilon$ are security parameters.

\vspace{1em}
\paragraph{\textbf{Secret Key Generation by DO:}} 
\begin{itemize}
    \item $\boldsymbol{s}=(s_1, s_2, \cdots ,s_{d+1}) \in \mathbb{R}^{d+1}$ and $\boldsymbol{\tau} \in \mathbb{R}^c$ are fixed, long-term secrets.
    \item $\boldsymbol{v}_i \in \mathbb{R}^{\epsilon}$  is a per-tuple ephemeral secret while  $\boldsymbol{r}^{(q)} \in \mathbb{R}^{c}$ and $\beta_q \in \mathbb{R}$ are per-query ephemeral secrets. 
    \item $\boldsymbol{M}$ is an invertible matrix with $(d+1+c+\epsilon)$ rows/columns and with elements drawn uniformly at random from  $\mathbb{R}$. 
    \item $\pi$ is a secret permutation function applied on $(d+1+c+\epsilon)$ elements. 
    \item The resultant secret key is $(\boldsymbol{M},\boldsymbol{s},\boldsymbol{\tau},\pi)$. 
\end{itemize}
\vspace{1em}
\paragraph{\textbf{Data Encryption:}} 
\begin{itemize}
    \item Each element of $\boldsymbol{p}_i$ is shifted by a value dictated by $\boldsymbol{s}$, and the point is augmented by the two secrets $\boldsymbol{\tau}$ and $\boldsymbol{v}_i$. DO therefore creates the tuple, 
    $$\hat{\boldsymbol{p}_i}=(s_1- 2p_{i_1}, \cdots, s_{d}-2p_{i_d},  s_{d+1}+ \Vert \boldsymbol{p}_i \Vert^2, \boldsymbol{\tau} ,  \boldsymbol{v}_i)$$. 
\item DO then computes the ciphertext,
$$\boldsymbol{p}_i'=\hat{\boldsymbol{p}}_i \hat{\boldsymbol{M}}^{-1}$$ 
Here, $\hat{\boldsymbol{M}}= \pi \boldsymbol{(M)}$, is the matrix obtained by permuting the columns of $\boldsymbol{M}$ using the permutation function, $\pi$ .   \cite{ZhuHuang} applies  $\pi$ on $\hat{\boldsymbol{p}}_i$ . However, permuting $\boldsymbol{M}$ instead of $\hat{\boldsymbol{p}}_i$ has the same effect on $\boldsymbol{p}_i'$ and simplifies the understanding of the encryption algorithm.
\end{itemize}

\paragraph{\textbf{Query Encryption:}} This proceeds as a protocol between the QU and DO.\\

The QU here uses Pallier encryption to encrypt the query point. A Paillier public key/private key pair is first generated. The details of this encryption scheme may be found in \cite{Paillier}. For purposes of this work, however, the following two properties of Pallier encryption are relevant: 
\begin{itemize}
\item $E_{pk}(m_1) \times E_{pk}(m_2) = E_{pk}(m_1+m_2) $  Property A
\item $ (E_{pk}(m_1))^c = E_{pk}(c \times m_1) $          \hspace{4.3em} 	Property B
\end{itemize}
Here, $pk$ is the Paillier public key, $m_1$ and $m_2$ are messages, and $c$ is a positive integer.\\

Having established this, the Query Encryption protocol is as follows:
\begin{itemize}
    \item Let $\boldsymbol{q}_{*}=(q_{*1}, q_{*2},\cdots,q_{*d})$ be the query issued. The QU encrypts each element using $pk$ and sends $E_{pk}(\boldsymbol{q}_*)=(E_{pk}(q_{*1}), \cdots,E_{pk}(q_{*d}))$ with $pk$ to the DO. 
    \item The elements of $\boldsymbol{q}_*$ have already been encrypted by the QU, thus preserving the privacy of the query against DO.
    \item DO generates ephemeral secrets, $\boldsymbol{r}^{(q)}$ and $\beta_q$. 

    \item It constructs the tuple, $(E_{pk}(\boldsymbol{q}_*), 1, \boldsymbol{r}^{(q)}, \boldsymbol{0}_{(\epsilon)})$. Elements of $\boldsymbol{q}_*$ have been encrypted by QU, thus the remaining $(c+\epsilon+1)$ elements are each Paillier-encrypted by the DO using  $pk$ as $$E_{pk}(\hat{\boldsymbol{q}}_*)=(E_{pk}(\boldsymbol{q}_*), E_{pk}(1), E_{pk}(\boldsymbol{r}^{(q)}), E_{pk}(\boldsymbol{0}_{(\epsilon)}))$$.
    
    \item The DO then computes a tuple of $(d+1+c+\epsilon)$ elements - the $i^{th}$ element of which is,  
\begin{align*}
\prod_{j=1}^{d+1+c+\epsilon} E_{pk}(\hat{q}_{*j})^{\beta_q \hat{m}_{ij}} \\
 & \hspace{-9em} =\prod_{j=1}^{d+1+c+\epsilon} E_{pk}(\beta_q \hat{m}_{ij} \times \hat{q}_{*j}) \hspace{5em}  \text{Property\ B} \\
 & \hspace{-9em} =E_{pk}\big( \sum_{j=1}^{d+1+c+\epsilon} (\beta_q \hat{m}_{ij} \times \hat{q}_{*j})\big) \hspace{4em} \text{Property\ A}
\end{align*}
The doubly-encrypted query tuple is sent by DO to QU.

\item QU then Pallier-decrypts each element of the received tuple using $sk$. Upon decryption, the resulting tuple is now encrypted with only the DO’s secret key and the $i^{th}$ element of the query tuple is of the form,
$$\sum_{j=1}^{d+1+c+\epsilon} \beta_q.\hat{m}_{ij} \times \hat{q}_{*j}$$ 
We denote encrypted query tuple by $\boldsymbol{q}_{*}'$ and this is sent to the CSP for the $k$-nearest neighbours computation
$$\boldsymbol{q}_{*}'=\beta_q.\hat{\boldsymbol{M}} \hat{\boldsymbol{q}}_*$$
 \item The query point $\boldsymbol{q}_{*}'$ is encrypted by DO's secret key, thus it holds the \textit{Query Privacy against CSP} of the encryption scheme.
 
\end{itemize}

\paragraph{\textbf{Query Processing:}}
\label{Query_process}
In its simplest form, solving the $k$-nearest neighbours problem involves computing the distance, $D(\boldsymbol{p}_i, \boldsymbol{q})$ between the query point, $\boldsymbol{q}$ and each  database point, $\boldsymbol{p}_i$. However, in the current context, we are dealing with, both, an encrypted query point and an encrypted database. Fortunately, a comparison between the distances, $D(\boldsymbol{p}_i, \boldsymbol{q})$ and $D(\boldsymbol{p}_j, \boldsymbol{q})$ reduces to a simple comparison in the encrypted domain as derived below:
\begin{align*}
D(\boldsymbol{p}_i, \boldsymbol{q}) > D(\boldsymbol{p}_j, \boldsymbol{q}) \\
& \hspace{-9em} \Leftrightarrow \Vert \boldsymbol{p}_i-\boldsymbol{q} \Vert^2 > \Vert \boldsymbol{p}_j-\boldsymbol{q}\Vert^2 \\
& \hspace{-9em} \Leftrightarrow \Vert \boldsymbol{p}_i \Vert^2-2\boldsymbol{p}_i\boldsymbol{q} + \Vert \boldsymbol{q} \Vert^2 > \Vert \boldsymbol{p}_j \Vert^2-2\boldsymbol{p}_j\boldsymbol{q} + \Vert \boldsymbol{q} \Vert^2 \\
& \hspace{-9em} \Leftrightarrow -2\sum_{k=1}^d p_{ik}q_k +  \Vert \boldsymbol{p}_i \Vert^2 > -2\sum_{k=1}^d p_{jk}q_k +  \Vert \boldsymbol{p}_j \Vert^2 \\
& \hspace{-9em} \Leftrightarrow  \sum_{k=1}^d (s_k- 2p_{ik})q_k + s_{d+1} + \Vert \boldsymbol{p}_i \Vert^2+ \boldsymbol{\tau}.\boldsymbol{r}^{(q)} \\
& \hspace{-9em} \hspace{0.5cm} > \sum_{k=1}^d (s_k- 2p_{jk})q_k + s_{d+1} + \Vert \boldsymbol{p}_j\Vert^2 + \boldsymbol{\tau}.\boldsymbol{r}^{(q)} \\
& \hspace{-9em} \Leftrightarrow \boldsymbol{p}_i' \boldsymbol{q}' >  \boldsymbol{p}_j' \boldsymbol{q}' 
\end{align*}
\section{Attacks}
\label{attack_mod}
In this section, we start by presenting a way to recover the columns of the secret key matrix $\hat{\mathbf{M}}$ and point $\boldsymbol{s}$; breaking \textit{key confidentiality} claim of the encryption scheme. The recovered information can be further used to break the \textit{data privacy} claim of the encryption scheme \cite{ZhuHuang}, and \cite{SinghKaul}, provided availability of the encrypted database. To break the \textit{Query Privacy} claim of the encryption scheme, the attacker requires to eavesdrop on the queries submitted by different query users to the CSP, followed by a few computations. 
The entire procedure is discussed in more detail below.


\subsection{Attack on key confidentiality and data privacy} 
\label{Attack_kc_dp}
The goal of the attacker is to retrieve the secret key matrix $\hat{\mathbf{M}}$ and use it to recover the database $\boldsymbol{\mathcal{D}}$ from $\boldsymbol{\mathcal{D}}'$. The attacker is allowed to execute $polynomial$-time cryptanalysis algorithms with respect to the dimension size of the database. Also, the attacker may have access to additional knowledge about the original data points, and the objective of the encryption algorithm is to prevent the attacker from obtaining $\boldsymbol{\mathcal{D}}$. Here, based on the knowledge the attacker can possess, we classify the attack scenarios as: 

\begin{itemize}
\item \textbf{\textit{Level 1}}: The attacker observes only $\boldsymbol{\mathcal{D}}'$, $\boldsymbol{q}'$ and the encrypted query results. This corresponds to the Ciphertext-Only Attack (COA) in cryptography \cite{ChenLiu}. 
\vspace{0em}
\item \textbf{\textit{Level 2}}: In addition to the information available in Level 1, the attacker also possesses some plaintext data  but he does not know which among $\boldsymbol{\mathcal{D}}'$ are the corresponding encrypted values. This corresponds to the known-sample attack in database cyptography \cite{Xu_Huiqi_Shumin_Guo}.
\vspace{1em}
\item \textbf{\textit{Level 3}}: In level 3 attack scenario the attacker have knowledge about the encrypted tuple $T'$ corresponding to plaintext tuple $T$ in addition to information available in Level 1. This is equivalent to known-plaintext attack (KPA) in cryptography \cite{Delfs}.
\end{itemize}

In this paper, we show that a Level 1 attack scenario can reveal certain essential information about the DO's secret key $\hat{\mathbf{M}}$ to the query user. This information can be leveraged in a Level 2 attack scenario to reveal long-term secret point $\boldsymbol{s}$, hence invalidating \textit{key confidentiality} claim of the encryption scheme. The recovered secret information can be used to recover $\boldsymbol{\mathcal{D}}$ from $\boldsymbol{\mathcal{D}'}$ provided the availability of $\boldsymbol{\mathcal{D}'}$ to the attacker. Thus, completely breaking the \textit{data privacy} claim of the encryption scheme.\\



A Query user following the encryption scheme \cite{ZhuHuang} for query encryption,  
the following is possible:
\begin{enumerate}
    \item From the attack on query controllability in \cite{SinghKaul}, given $\boldsymbol{q}'=\beta_q.\hat{\mathbf{M}}.\boldsymbol{\hat{q}}^{T}$, QU can find $\beta_q$.
    This can be extended to extract all columns of the secret $\hat{\mathbf{M}}$ that gets multiplied to data point co-ordinates. 
    \item Let $N$ be an integer of large magnitude. QU formulate a  query point $\boldsymbol{q}_j =e_j.N =(0, \cdots, 0, N, 0, \cdots, 0)$. Here, $e_j$ representing a point of length same as that of $\boldsymbol{q}'$  with all zeros except the $j$-th position.
    \item After the query encryption protocol with DO, QU gets $\boldsymbol{q}'_j=\beta_q.\hat{\mathbf{M}}.\boldsymbol{\hat{q}}_j^{T}$ corresponding to $\boldsymbol{q}_j$. He removes $\beta_q$ as mentioned in \cite{SinghKaul}. Let this be $\boldsymbol{t}$
    \item $\boldsymbol{t}$ is therefore a point such that for each element 
    $$t_i = \boldsymbol{\hat{M}}_{ij}\cdot N + \boldsymbol{\hat{M}}_{ik} + \boldsymbol{\hat{M}}_{i*}\cdot \boldsymbol{r}^{(q)}$$ 
    where index $j$ gets multiplied to $N$, index $k$ gets multiplied to $1$ and indices $i*$ get multiplied to different elements of point $\boldsymbol{r}^{(q)}$. 
    \item Since $N$ is a large number, the following can be derived:
    $$\text{column }\ \boldsymbol{\hat{M}}_{j} = \frac{\boldsymbol{t} - (\boldsymbol{t} \ mod \ N)}{N}$$
    \item $d$, similar queries can be used to get all $d$-columns of $\mathbf{\hat{M}}$ that correspond to the columns multiplied with elements in $\boldsymbol{q}_*$. 
\end{enumerate}
To attack the encryption scheme in \cite{SinghKaul}, we need an assumption that the attacker posing as QU resides in the CSP (very much possible in practical scenarios).
In such a scenario, $\mathbf{W}$ is known, and so $\boldsymbol{q}'$ can be derived. However, $\boldsymbol{q}'=\mathbf{\hat{M}}.\boldsymbol{\hat{q}}^{T}$ and $\beta_q$ is only multiplied to certain elements in $\boldsymbol{\hat{q}}$, the indices of which are unknown. Nevertheless, it suffices to find the GCD of $n+1$-subsets of positions and check when the answer that is not 1 is found. That would be the value of $\beta_q$. The rest of the steps are the same as that of mention earlier.\\

Retrieval of 
the long-term secret point, $\boldsymbol{s}$, paves the way to retrieval of the entire database of tuples in the clear. For this purpose, we employ a strategy akin to a known sample attack, albeit with a more restrictive assumption.  
Considering \textit{Level 2 attack} scenario, we have access to the entire database of encrypted tuples together with the tuple-to-index mappings.  In addition, we know a small number of tuples in the clear. 

If even a single tuple, $\boldsymbol{p}_i$ and its corresponding ciphertext, $\boldsymbol{p}_i'$ are known together with selected columns of $\boldsymbol{\hat{M}}$ (obtained from \textit{Level 1 attack}),  the entire secret point, $\boldsymbol{s}$, can be computed. Since $\boldsymbol{p}_i'.\boldsymbol{\hat{M}} = \hat{\boldsymbol{p}}_i$ and from the definition of  $\boldsymbol{\hat{p}}_i$ ,   we get
\begin{align}
s_j=2p_{ij} + < \boldsymbol{p}_i'\ , \pi(\boldsymbol{M})>
\label{attack_eq}
\end{align}

However, we do not know which of the encrypted tuples in the database are the ciphertexts of our known-plaintext tuples. Our attack thus differs from the vanilla known-plaintext attack since the latter typically assumes that, in addition to the plaintexts, we also know the corresponding ciphertexts. 

\begin{table*}[htbp]
  \centering
  \begin{tabular}{|c|c|c|c|c|}
    \hline
    \multicolumn{1}{|p{2cm}|}{Encrypted Tuples in Database} &\multicolumn{4}{c|}{Known Plaintexts} \\
     \cline{2-5}
    & $\makecell{(\boldsymbol{p}_{i_{1}}) \vspace{.5em} \\  s_1' s_2' \cdots s_d'}$ & $\makecell{(\boldsymbol{p}_{i_{2}}) \vspace{.5em} \\ s_1' s_2' \cdots s_d'}$ & $\cdots$ & $\makecell{(\boldsymbol{p}_{i_{k}}) \vspace{.5em} \\  s_1' s_2' \cdots s_d'}$ \\
    
    \cline{1-5}    
    $\boldsymbol{p}_1'$ & $123 \ 939 \cdots 478$ & $571 \ 286 \cdots 329$ & $\cdots$ & $295 \ 310 \cdots 836$ \\
    \cline{1-5}
     $\boldsymbol{p}_2'$ & $752 \ 083 \cdots 906$ & $472 \ 388 \cdots 512$ & $ \ \ \ \ \ \ \ \ \cdots \ \ \ \ \ \ \ \ $ & $185 \ 364 \cdots 280$ \\
    \cline{1-5}
     $\vdots$ & $\vdots$ & $\vdots$ & $\vdots$ & $\vdots$ \\
    \cline{1-5}
     $\boldsymbol{p}_t'$ & $295 \ 310 \cdots 836$ & $065 \ 387 \cdots 926$ & $\cdots$ & $693 \ 377 \cdots 598$\\
   \hline  
  \end{tabular}
    \vspace{1em}
    \caption{Table aiding the attack technique} 
    \vspace{-.5cm}
    \label{Attack_Tab}
\end{table*}

Given this information,  we construct a table as shown in the table \ref{Attack_Tab}. The table presents an example listing with the encrypted tuples in the leftmost column. Each of the $k$ remaining columns corresponds to a known plaintext. A cell shows the presumed key, $\boldsymbol{s}’$, computed by Equation \ref{attack_eq} for an encrypted tuple, known-plaintext pair.
   
The pseudocode used to populate Table \ref{Attack_Tab} is shown below.

\begin{algorithm}
 \KwData{$\boldsymbol{p}^{'}_{r},\boldsymbol{\hat{M}},\boldsymbol{p}_{i}$}
 \KwResult{$s_1,\cdots,s_d$}
 \For{$(j=1,\cdots,k)$}{  \tcp{for each known plaintext, $\boldsymbol{p}_{i_j}$}
    \For{$(r=1,\cdots,t)$}{  \tcp{for each encrypted tuple in the DB, $\boldsymbol{p}_r'$}
        \For{$(u=1,\cdots,d)$}{  \tcp{for each element of the secret point, $\boldsymbol{s}'$}
            $s_u'=2p_{i_ju} + <\boldsymbol{p}_r',\pi(\boldsymbol{M})>$ 
        }
    }
}
 \caption{Pseudocode to populate table \ref{Attack_Tab}}
\end{algorithm}

The true value of $\boldsymbol{s}$ must appear in each of the $k$ columns, but, as we show next, there is a negligible probability of an incorrect value appearing in more than a single column. 

Estimating the probability of any given value of $\boldsymbol{s}’$ appearing in two columns is analogous to computing the probability of being able to break the weak collision resistance of a cryptographic hash function. Given a hash function which returns one of $n$ distinct values and given $m$ different messages, $m << n$, the probability of a hash collision is  $(1- e^{-\frac{m^2}{n}})$. This is an upper bound on the probability of a hash collision between two lists of hash values each of size  $\frac{m}{2}$.

Here, let $t$ denote the number of encrypted tuples in the database. So $m=2t$. For $d$-dimensional tuples and $w$ decimal digits per element of the secret point , $n=10^{wd}$. So, the probability of an incorrect secret appearing in two columns is less than   $1- e^{-\frac{4t^2}{10^{wd}}}$.  Assuming a database of $10^6$,  $20$-dimensional tuples and $3$ significant digits per element of the secret key, $\boldsymbol{s}$, the above probability evaluates to  $(1- e^{-4\times 10^{-48}}) \approx 0$ . Thus, even with as few as two known plaintexts, the probability of an incorrectly guessed secret repeating across columns is vanishingly small. On the other hand, the true secret will necessarily repeat across all columns and so be easily identified. 

Once $\boldsymbol{s}$ is known, the plaintext of each encrypted tuple may be obtained using Equation \ref{attack_eq}, breaking \textit{data privacy} claim of the encryption scheme.\\
\vspace{-2em}
\subsection{Attack on Query privacy:} 
The goal of this section is to show that an attacker having access to the information retrieved in the section \ref{Attack_kc_dp} can recover 
the plaintext query point given an encrypted query point.
By snooping on the communication link between QU and CSP 
the attacker obtains  $\boldsymbol{q}_*'$  (the encrypted query, encrypted using the DO's secret key).  From this, we show that he can compute  $\boldsymbol{q}_*$ ,  the plaintext form of the encrypted query point.

Given $\hat{\boldsymbol{p}}_i=(s_1- p_{i_1}, \cdots ,  s_d- p_{i_d}, s_{d}+ \Vert \boldsymbol{p}_i\Vert^2, \boldsymbol{\tau}, \boldsymbol{v}_i)$,  $\hat{\boldsymbol{q}}_*=(\boldsymbol{q}_* , 1 , \boldsymbol{r}^{q} , \boldsymbol{0}_{\epsilon} )$, and $\hat{\boldsymbol{M}}^{-1}= (\pi(\boldsymbol{M}))^{-1}$, 
we have 
\begin{align}
\boldsymbol{p}_i'= \hat{\boldsymbol{p}}_i \hat{\boldsymbol{M}}^{-1}    
\label{enc_plain}
\end{align}                                                         
\begin{align}
\boldsymbol{q}_*^{'T}= \beta_q\hat{\boldsymbol{M}}\hat{\boldsymbol{q}}_*^{T}    
\label{enc_query}
\end{align}

From Equations \ref{enc_plain} and \ref{enc_query} , it follows that
\begin{align}
\hat{\boldsymbol{p}}_i\hat{\boldsymbol{q}}_*^T= \boldsymbol{p}_i'\Big(\frac{\boldsymbol{q}_*^{'T}}{\beta}\Big) \label{plain_cipher_mul}    
\end{align}

In addition to the encrypted query,  $\boldsymbol{q}_*'$, we have a copious collection of (plaintext, ciphertext) pairs,  $(\boldsymbol{p}_{i_1}, \boldsymbol{p}_{i_1}'), (\boldsymbol{p}_{i_2},\boldsymbol{p}_{i_2}'), \cdots ,(\boldsymbol{p}_{i_k}, \boldsymbol{p}_{i_k}')$ obtained from previous attacks. Expanding Equation \ref{plain_cipher_mul} for each (plaintext, ciphertext) pair yields equations as that of figure 1.

\begin{figure*}[!t]
\normalsize
\hrulefill

\begin{align*}
    (s_1- 2p_{i_{1}1})q_{*1} + \cdots + (s_d- 2p_{i_1d})q_{*d} + s_{d+1} + \Vert \boldsymbol{p}_{i_1} \Vert^2 + <\boldsymbol{\tau},\boldsymbol{r}^{(q)}> &= <\boldsymbol{p}_{i_{1}}' , \Big(\frac{\boldsymbol{q}_*^{'}}{\beta}\Big)^{T}>
\end{align*}
\begin{align*}
    (s_1- 2p_{i_{2}1})q_{*1} + \cdots + (s_d- 2p_{i_{2}d})q_{*d} + s_{d+1} + \Vert \boldsymbol{p}_{i_2} \Vert^2 + <\boldsymbol{\tau},\boldsymbol{r}^{(q)}> &= <\boldsymbol{p}_{i_{2}}' , \Big(\frac{\boldsymbol{q}_*^{'}}{\beta}\Big)^{T}>
\end{align*}
    \hspace{15em}  \vdots 
\begin{align*}
    (s_1- 2p_{i_{k}1})q_{*1} + \cdots + (s_d- 2p_{i_{k}d})q_{*d} + s_{d+1} + \Vert \boldsymbol{p}_{i_k} \Vert^2 + <\boldsymbol{\tau},\boldsymbol{r}^{(q)}> &= <\boldsymbol{p}_{i_{k}}' , \Big(\frac{\boldsymbol{q}_*^{'}}{\beta}\Big)^{T}>
\end{align*}
    
    Subtracting first equation from rest of the equations, and rearranging the terms, we solve the unknown query as:
\begin{equation*}
        \begin{pmatrix}
        q_{*1} \\ q_{*2} \\ \vdots \\ q_{*d} \end{pmatrix}= 
        \begin{pmatrix}
        2(p_{i_{1}1}- p_{i_{2}1}) \ \cdots \ 2(p_{i_{1}d} - p_{i_{2}d}) \\
        2(p_{i_{1}1} - p_{i_{3}1}) \ \cdots \ 2(p_{i_{1}d}- p_{i_{3} d}) \\
        \vdots  \\ 
        2(p_{i_{1}1} - p_{i_{k}1}) \ \cdots \ 2(p_{i_{1}d}- p_{i_{k} d})
        \end{pmatrix}^{-1}
        \begin{pmatrix}
        \Vert \boldsymbol{p}_{i_1} \Vert^2 - \Vert \boldsymbol{p}_{i_2} \Vert^2-  <(\boldsymbol{p}_{i_{1}}'-\boldsymbol{p}_{i_{2}}'),\Big(\frac{\boldsymbol{q}_*^{'}}{\beta}\Big)^{T}> \\
        \Vert \boldsymbol{p}_{i_1} \Vert^2 - \Vert \boldsymbol{p}_{i_3} \Vert^2-  <(\boldsymbol{p}_{i_{1}}'-\boldsymbol{p}_{i_{3}}'),\Big(\frac{\boldsymbol{q}_*^{'}}{\beta}\Big)^{T}> \\
        \vdots \\ 
        \Vert \boldsymbol{p}_{i_1} \Vert^2 - \Vert \boldsymbol{p}_{i_k} \Vert^2-  <(\boldsymbol{p}_{i_{1}}'-\boldsymbol{p}_{i_{k}}'),\Big(\frac{\boldsymbol{q}_*^{'}}{\beta}\Big)^{T}>
        \end{pmatrix}
\end{equation*}
\hrulefill
\label{Attack_query_Privacy}
\caption{Expansion of equations for attacking query privacy of the encryption scheme}
\end{figure*}

\normalsize
We choose a total of $k = d+1$  independent (plaintext, ciphertext) pairs for this attack. All terms on the RHS are known or may be computed. (Note also that $\beta_q$ may be computed as explained in \cite{SinghKaul}).  If the choice of (plaintext, ciphertext) pairs results in the above matrix being singular, we repeat the computation with different choices until a non-singular matrix is obtained. This completes the second attack designed to eavesdrop on the queries submitted by different clients to the CSP.

\subsection{Known plaintext attack on query privacy and data privacy:} 
\label{Known_Plaintext_Attack}
The encryption scheme uses the model of exact Euclidean distance comparison for finding k-NN in the CSP. Thus a curious cloud having access to the polynomial number of independent (plaintext, ciphertext) pairs in addition to the encrypted data points and encrypted query points asked by the querying user can recover the plaintext query point. After recovery of the polynomial number of plaintext query points, he can further use it to recover the plaintext data points breaking \textit{data privacy} claim of the encryption scheme. More details about the attack can also be found in \cite{LinWeipeng}.

\section{Proposed Encryption Scheme}
\label{Proposed_scheme}
We propose an enhanced encryption scheme that provides different features, viz. \textit{Data Privacy}, \textit{Query Privacy}, and \textit{Key Confidentiality}. \textit{Query Controllability} and \textit{Query Check Verification} in the enhanced encryption scheme can be easily obtained; thus, we are not going to cover it here. In the proposed encryption scheme to encrypt a query point, random variables are used, which are generated as a function of the query point, unlike earlier encryption schemes. Use of random variables as a function of query point for query encryption results in mitigation of attacks on \textit{key confidentiality} discussed above. We will cover it in more detail in the security analysis section below.

The proposed encryption scheme can even withstand an attack on $\textit{Query Privacy}$ by a curious CSP having access to a polynomial number of plaintext ciphertext pairs as mentioned by Lin et al.\cite{LinWeipeng}. The encryption scheme uses an extra random variable for tuple encryption and, accordingly, for query encryption to obtain a relative distance comparison for $k$-NN computation. Thus, preserving the encryption scheme's $\textit{Query Privacy}$.
The different modules of the proposed encryption scheme are

\begin{enumerate}
\item \textbf{KeyGen:} DO generates the following:
    \begin{itemize}
        \item positive integer $\epsilon$ and $c$ as security parameters with $1<c<d$.
        \item invertable matrix $\mathbf{M}$ $\in {R}^{\eta \times \eta}$ where $\eta=d+2+c+\epsilon$
        \item $(d+1)$-dimensional shift point $\boldsymbol{s}=(s_1, s_2, s_3, \cdots , s_{d+1}) \in \mathbb{R}^{(d+1)}$
        \item $d$-dimensional long term secret $\boldsymbol{\sigma}=(\sigma_1, \sigma_2, \sigma_3, \cdots , \sigma_{d}) \in \mathbb{R}^{d}$, Sampled in such a way that each tuple of the point $\boldsymbol{\sigma}$ is greater than all plaintext tuple corresponding to that index.
        \item $(c+\epsilon)$-dimensional bit string $\boldsymbol{b}=(b_1,b_2,b_3,\cdots,b_{c+\epsilon}) \in \{0,1\}^{c+\epsilon}$ sampled with initial $c$ bits as 1's and remaining $\epsilon$ random bits with at least one 0 and one 1 bit.
        \item $(c+\epsilon)$-dimensional weight point $\boldsymbol{w} =(w_1,w_2,w_3, \cdots , w_{c+\epsilon})\in \mathbb{R}^{c+\epsilon}$.
        \item permutation function $\boldsymbol{\pi}$ of $\eta$ positions
    

    \end{itemize}
    
    The secret key of the DO is $(\boldsymbol{\mathbf{M}},\boldsymbol{s},\boldsymbol{\sigma},\boldsymbol{b},\boldsymbol{w},\boldsymbol{\pi})$.\vspace{1em}
    
    In the proposed encryption scheme, the content of weight point $\boldsymbol{w}$ is used in both tuple encryption and query encryption in combination with randomly generated real numbers based on bit-vector $\boldsymbol{b}$. So that the final product of artificially added variables used in tuple encryption and query encryption results in zero\cite{WongCheung}.
    
    \item \textbf{Tuple Encryption:} To encrypt a $d$-dimensional data point $\boldsymbol{p}_i \in \boldsymbol{D}$ following steps need to be followed:
    \begin{enumerate}
        \item[1 -] compute a $(c+\epsilon)$-dimensional ephemeral secret point $\boldsymbol{\tau}_i$ s.t. for each element, 
        $$\tau_{ij} =
            \begin{cases}
                w_j    & \text{if } b_j = 1 \\
                rand   & \text{if } b_j = 0 \text{ but not the last 0}\\

                -nom_p    & \text{otherwise}
            \end{cases}
        $$
            
        here $rand$ is a random real number generated uniformly at random for each $b_j=0$ (except the last) and $nom_p$ is a normalizer such that $\boldsymbol{w}.\boldsymbol{\tau}_i = 0$   
        \item[3 -] create $\hat{\boldsymbol{p}_i} = (s_1-2p_{i1},s_2-2p_{i2},\cdots,s_d-2p_{id}, s_{d+1}+\Vert \boldsymbol{p}_i \Vert^{2}, \alpha_i, \boldsymbol{\tau}_i)$ \\
        Where $\alpha_i$ is computed as
         $$\alpha_i=(\boldsymbol{\sigma}_{max}-t_i)^2$$ 
         with positive $t_i$ and $t_i<\boldsymbol{\sigma}_{max}$, selected uniformly at random for each data point encryption.
        \item[4 -] encrypt it as $\boldsymbol{p}_i' = \hat{\boldsymbol{p}}_i\mathbf{\hat{M}^{-1}}$.
        Here, $\hat{\boldsymbol{M}}= \boldsymbol{\pi} \boldsymbol{(M)}$. 
    \end{enumerate}
    \vspace{1em}

\item \textbf{Query Encryption:} Query encryption involves exchange of following messages between QU and DO:
    \begin{enumerate}
        \item[1 -] QU generates key pair $(pk,sk)$ of Paillier encryption scheme.
        \item[2 -] encrypt query point $\boldsymbol{q}$ to obtain $\dot{\boldsymbol{q}} = (E_{pk}(q_1), E_{pk}(q_2), \cdots, E_{pk}(q_d))$. 
        \item[3 -] send ($\dot{\boldsymbol{q}}$ , $pk$) to DO.
        \item[4 -] if DO decides to not allow the query, it returns $\perp$ to QU. Otherwise, it does the following:
        \begin{itemize}
            \item select a random real number $\beta_1 \in \mathbb{R}^{+}$
            \item select a real number $\beta_2  > \beta_1 \cdot |\boldsymbol{\sigma}_{max}|^2$
            \item create a point $\boldsymbol{v}$, a random permutation of $\{0,1,2,\cdots,d-1\}$ 
            \item compute $\boldsymbol{r}^{enc}$, a $(c+\epsilon)$-dimensional point as shown in the Figure 2. 

\begin{figure*}[!t]
\normalsize


\hrulefill
 $$r^{enc}_j =
            \begin{cases}
                (-\prod_{t=j*\lfloor \frac{d}{c} \rfloor}^{(j+1)*\lfloor \frac{d}{c} \rfloor-1}(\dot{q}_{v[t]}))^{rand}    & \text{0 $\leq j < c-1$} \\
                (-\prod_{t=j*\lfloor \frac{d}{c} \rfloor}^{d}(\dot{q}_{v[t]}))^{rand}    & \text{$j=c-1$} \\
                -E_{pk}(w_j)    & \text{if } b_j = 0 \\
                -E_{pk}(rand) & \text{if $c \leq j<(c+\epsilon)$}, b_j = 1 \text{ (not last 1)} \\
                E_{pk}(nom_q)    & \text{otherwise}
            \end{cases}$$

\hrulefill
\label{r_q_value}
\caption{Computation of $r^{enc}$ value in the proposed encryption scheme}
\end{figure*}
            
        In the Figure $rand$ represents a random number generated uniformly at random for each $b_j=1$ (except the last 1) and $nom_q$ is a normalizer such that  $\boldsymbol{r}^{enc}.\boldsymbol{w} = E_{pk}(0)$.
            \item compute \\ $\overline{\boldsymbol{q}}=(\boldsymbol{\hat{q}},E_{pk}(\beta_2), E_{pk}(\beta_1), \boldsymbol{r}^{enc})$  where  $d$-dimensional $\hat{\boldsymbol{q}}$ is obtained by computing $\hat{q_{i}} = \dot{q}_{i}^{\beta_2}$ for each dimension. 
            \item calculate $\mathbf{{a}}^{(q)}=(a_1^{(q)},a_2^{(q)},\cdots,a_{\eta}^{(q)})$ as $a_{i}^{(q)}=\prod_{j=1}^{\eta} (\overline{\boldsymbol{q}}_j)^{(\boldsymbol{\hat{M}}_{i,j})}$ 

            
            \item return $\mathbf{{a}}^{(q)}$ to QU
        \end{itemize}
        \item[5 -] QU decrypts each component of obtained $\mathbf{{a}}^{(q)}$ using $sk$ to get $\boldsymbol{q}'$ i.e.
        \vspace{1em}\\
            $q'_i=D_{sk}(a_i^{(q)})$    
        \vspace{1em}
        \item[6 -] QU sends $\boldsymbol{q}'$ to CSP, where 
        \begin{align*}
        \boldsymbol{q}' &=  \mathbf{\hat{M}}.(\boldsymbol{q}\beta_2,\beta_2,\beta_1,\boldsymbol{r}^{dec}) \\
        \boldsymbol{r}^{dec} &=D_{sk}(\boldsymbol{r}^{enc})
        \label{qp_point1}
        \end{align*}
    \end{enumerate}
     \item \textbf{Secure $k$-NN Computation:} $k$-NN computation using the proposed encryption scheme can be computed in the same way as that of \cite{SinghKaul}. The proposed encryption scheme is framed in such a way that dot product between the added randomness $\boldsymbol{\tau}$ and $\boldsymbol{r}^{enc}$ results in $\Vert \boldsymbol{w} \Vert^{2}$. Thus, when the distance between a query point from data points is compared, $\Vert \boldsymbol{w} \Vert^{2}$ gets canceled out, resulting in the correct $k$-NN value.
    
    
    \item \textbf{Tuple Decryption:}  For an encrypted data point $\boldsymbol{p}_i'$, 
    , $\hat{\boldsymbol{p}_i}=\boldsymbol{p}_i'\mathbf{\hat{M}}$. $\boldsymbol{p}_i$ can be extracted from $\hat{\boldsymbol{p}_i}$ by $\boldsymbol{p}_i=\frac{\boldsymbol{s}-\hat{\boldsymbol{p}}_i}{2}$. 
\end{enumerate}

\begin{Claim}
\label{tau_q_proof}
Multiplication of $\boldsymbol{\tau}_k$ and  $\boldsymbol{r}^{dec}=D_{sk}(\boldsymbol{r}^{enc})$ evaluates to $\sum_{i=1}^{c+\epsilon} w_i^2=||\boldsymbol{w}||^2$; [$k$ any random index]
\end{Claim} 

\paragraph{Proof:}
From the definition $nom_p$ in $\boldsymbol{\tau}_k$ we have $\boldsymbol{w}.\boldsymbol{\tau}_k = 0$. Similarly, $nom_q$ in $\boldsymbol{r}^{enc}$ results in $\boldsymbol{w}.\boldsymbol{r}^{enc} = E_{pk}(0)$  ($\boldsymbol{w}.\boldsymbol{r}^{dec} = 0$ after decryption). Suffix $0/1$ in $w$ corresponds to $\boldsymbol{w}$ point content corresponding to $0/1$ element of bit point $\boldsymbol{b}$ in the respective position as presented in the Figure 3. 

\begin{figure*}[!t]
\normalsize


\hrulefill
\begin{align*}
    \boldsymbol{\tau}.\boldsymbol{r}^{dec} &= -w_1. rand.q -rand.w_1 + w_1.nom_q - rand.w_0  + nom_p.w_0 \\
    &= -w_1. rand.q -rand.w_1 + (\sum w_{1}^{2}+w_1.rand.q + rand.w_1) - rand.w_0  + (\sum w_{0}^{2} + rand.w_0) \\
    &[\ Expanding\ nom_q\ and\ nom_p]\\
    &=\sum w_{1}^{2} + \sum w_{0}^{2}\\ &=\sum_{i=0}^{c+\epsilon} w_i^2 \\ &=||\boldsymbol{w}||^2
\end{align*}

\hrulefill
\label{Claim}
\caption{proof of the claim \ref{tau_q_proof}}
\end{figure*}
\vspace{1em}

\subsection{Correctness Analysis}
Correctness of the encryption scheme needs two-fold proof. Firstly, it is necessary to show that the decryption of the encrypted data point under the scheme yields the correct value. This is evident as it directly follows from the \textit{tuple encryption} and \textit{tuple decryption} phases of the encryption scheme.

Next, we show that the encrypted query point $\boldsymbol{q}'$  can be used to find the correct nearest neighbors. Assuming encrypted query point $\boldsymbol{q}'$  is valid encryption of the query point $\boldsymbol{q}$, CSP tries to find the nearest neighbor from the encrypted database $\boldsymbol{\mathcal{D}'}$ by comparing the distance between database points. Correctness of $k$-NN calculation for the proposed encryption scheme can be summarized by the following lemma: 

\begin{lemma}
\label{Lemma_plain_cipher_same_dist}
Let $\boldsymbol{p}'_1$ and $\boldsymbol{p}'_2$ represent the encryption of plaintext point $\boldsymbol{p}_1$ and $\boldsymbol{p}_2$ respectively and $\boldsymbol{q}'$ represent the encryption of the query point $\boldsymbol{q}$. Then the homomorphic property of the proposed encryption scheme can correctly evaluate which of the two is the nearest neighbour.
\end{lemma}

\paragraph{Proof:} The proof for this follows from the fact that the encryption scheme preserves euclidean distance comparison in the encrypted domain, as shown in figure 4 below. 
\\

\begin{figure*}[!t]
\normalsize
\hrulefill
\begin{align*}
(\boldsymbol{p}'_1-\boldsymbol{p}'_2)\cdot\boldsymbol{q}'  
&= (\boldsymbol{\hat{p}}_{1}\cdot\boldsymbol{\hat{M}}^{-1} - \boldsymbol{\hat{p}}_{2}\cdot\boldsymbol{\hat{M}}^{-1})\cdot \boldsymbol{q}'\\
&= (\boldsymbol{\hat{p}}_{1} - \boldsymbol{\hat{p}}_{2})\cdot \widetilde{\boldsymbol{q}}  
\hspace{10em}  
\widetilde{\boldsymbol{q}}=(\beta_2.\boldsymbol{q},\beta_2,\beta_1,\boldsymbol{r}^{dec}) \\
\\
\boldsymbol{\hat{p}}_1\cdot\widetilde{\boldsymbol{q}}    
&=(s_1-2p_{11},\cdots,s_d-2p_{1d}, s_{(d+1)}+\Vert \boldsymbol{p}_1 \Vert^{2},\alpha_1 ,\boldsymbol{\tau}_1)\cdot (\beta_2q_1,\beta_2q_2,\cdots,\beta_2q_d,\beta_2,\beta_1,\boldsymbol{r}^{dec}) \\
&=\beta_2 q_1(s_1-2p_{11}) + \cdots + \beta_2 q_d(s_d-2p_{1d}) + \beta_2(s_{(d+1)} + \Vert \boldsymbol{p}_1 \Vert^{2}) + \alpha_1 \beta_1 + \boldsymbol{\tau}_1\cdot \boldsymbol{r}^{dec}\\ 
&=\beta_2 q_1(s_1-2p_{11}) + \cdots + \beta_2 q_d(s_d-2p_{1d}) + \beta_2(s_{(d+1)} + \Vert \boldsymbol{p}_1 \Vert^{2}) + \alpha_1 \beta_1 + ||\boldsymbol{w}||^2 &\text{[From claim \ref{tau_q_proof}]} \\
\text{Similarly,} \\
\boldsymbol{\hat{p}}_2\cdot \boldsymbol{\widetilde{q}}    
&=\beta_2 q_1(s_1-2p_{21}) + \cdots + \beta_2 q_d(s_d-2p_{2d}) + \beta_2(s_{(d+1)}+\Vert \boldsymbol{p}_2 \Vert^{2}) + \alpha_2\beta_1 +||\boldsymbol{w}||^2 \\ \\
(\boldsymbol{\hat{p}}_{1}-\boldsymbol{\hat{p}}_{2})\cdot \boldsymbol{\widetilde{q}}   
&= \beta_2 \left( \Vert \boldsymbol{p}_1\Vert^{2}-2p_{11}q_1 - \cdots -2p_{1d} q_d \right)  -\beta_2(\Vert \boldsymbol{p}_2\Vert^{2}-2p_{21} q_1 - \cdots -2p_{2d} q_d) + \beta_1(\alpha_1-\alpha_2)  \\
&= \beta_2 \left( (p_{11}-q_1)^2 +  \cdots + (p_{1d}-q_d)^2 \right) -\beta_2 \left( (p_{21}-q_1)^2 + \cdots + (p_{2d}-q_d)^2 \right) + \beta_1(\alpha_1-\alpha_2)  \\
&= \beta_2 \left[ ED(\boldsymbol{p}_{1}-\boldsymbol{q})^2 - ED(\boldsymbol{p}_{2}-\boldsymbol{q})^2 \right] + \beta_1(\alpha_1-\alpha_2)
\end{align*}    
Correct $k$-NN will be obtained provided following condition gets satisfied
$$ | \beta_2 \left[ ED(\boldsymbol{p}_{1}-\boldsymbol{q})^2 - ED(\boldsymbol{p}_{2}-\boldsymbol{q})^2 \right] | > |\beta_1(\alpha_1-\alpha_2)| $$
Which is taken care by the condition 
$$\beta_2  > \beta_1 \cdot |\boldsymbol{\sigma}_{max}|^2$$
Provided $|ED(\boldsymbol{p}_{1}-\boldsymbol{q})^2 - ED(\boldsymbol{p}_{2}-\boldsymbol{q})^2|$ is greater than $1$, which is not hard to get. This can be done by converting all data point tuples and query point tuples into integers. This completes the proof of the lemma.

\hrulefill
\label{lemma}
\caption{Proof of the lemma \ref{Lemma_plain_cipher_same_dist}}
\end{figure*}

\subsection{Security Analysis} 
Security of the encryption scheme mainly focuses on \textit{Key Confidentiality}, \textit{Data Privacy} and \textit{Query Privacy}. We can say the presented encryption scheme is secure if each of these is secure. For concreteness below, we have covered each of them separately.

\subsubsection{Data privacy} 
Considering a \textit{Level 2 attack} scenario, the attacker has access to the entire encrypted database, encrypted query point, query answer, and some plaintext data points. The goal of the encryption scheme is to restrict attackers from breaching data privacy as well as query privacy. 
The encrypted data point is of the form 
$$\boldsymbol{p}_i' = \hat{\boldsymbol{p}}_i\mathbf{\hat{M}}^{-1}$$
where 
\begin{align*}
\hat{\boldsymbol{p}}_i &= (s_1-2p_{i1},\cdots,s_d-2p_{id}, s_{(d+1)}+\Vert \boldsymbol{p}_{i} \Vert^{2}, \alpha_i ,\boldsymbol{\tau}_i)
\end{align*}

Here, $\alpha_i$ and $\boldsymbol{\tau}_i$ is data encryption ephemeral secrets and changes each time even if the same data point is encrypted for the second time -- resulting in the utterly indistinguishable ciphertext. Thus even if an attacker has apriori-knowledge about data samples and tries to perform independent component analysis to approximately reconstruct the original data -- perturbed and multiplied by the secret matrix. It is not possible without the knowledge of one-time ephemeral secret $\alpha_i$ and $\boldsymbol{\tau}_i$. 

The use of random number $\alpha_i$ and point $\boldsymbol{\tau}_i$ in $\hat{\boldsymbol{p}}_i$ and permutation function $\boldsymbol{\pi}$ in $\boldsymbol{M}$ reduces $\boldsymbol{p}_i'$ into pseudo-random. Thus, from pseudo-random $\boldsymbol{p}_i'$ it is nearly impossible to get any information about the encrypted plaintext points.

\subsubsection{Query privacy} Query privacy against DO comes from the Paillier crypto-system. Paillier crypto-system \cite{Paillier} is known to be semantically secure; thus, it guarantees query privacy against curious DO. 

Query privacy against the curious CSP comes from the fact that the encrypted query point obtained by the CSP has a form of equation \ref{qp_point}. 

\begin{equation}
\boldsymbol{q}'=\mathbf{\hat{M}}.(\boldsymbol{q}\beta_2,\beta_2,\beta_1,\boldsymbol{r}^{dec})
\label{qp_point}
\end{equation}

Asymmetric scalar product preserving encryption using a random invertible matrix and random numbers during encryption can resist level-2 attacks such as principal component analysis, signature linking attack, duplicate analysis, brute-force attack, etc. \cite{WongCheung}. Use of random $\beta_1$, $\beta_2$ and $\boldsymbol{r}^{dec}$ for each query encryption in addition to multiplication with $\boldsymbol{\hat{M}}$ completely hides the query $\boldsymbol{q}$.

Though the CSP have access to $\eta$-dimensional $query$ point $\boldsymbol{q}'$ calculated using equation \ref{qp_point}.
The information $\boldsymbol{q}'$ without the knowledge of $\beta_1$, $\beta_2$, $\boldsymbol{r}^{dec}$ and $\boldsymbol{\hat{M}}$ alone is not sufficient to frame equations to solve for plain query point $\boldsymbol{q}$. Thus, in short the proposed scheme can preserve query privacy against CSP under a level-2 attack scenario.

\subsubsection{Key Confidentiality:} Considering a \textit{Level 1 attack scenario}, the attacker knows the encrypted database, encrypted query point, and computed $k$-NN values. Furthermore, to learn the secret information of DO, the attacker can also possess as a QU.

In the query encryption phase, the attacker posing as QU can ask encryption of $\boldsymbol{q}$ of its choice. DO encrypts it and responds with $\mathbf{{a}}^{(q)}$. Attacker removes the Paillier encryption layer and recovers the encrypted query point $\boldsymbol{q}'$ as shown in equation \ref{qp_point}. Each element present in $\boldsymbol{q}'$ expands as shown in equation \ref{q_tilda}.

\begin{equation}
\hspace{-1em} q'_i= \beta_2 \left(\sum_{j=0}^{d} \boldsymbol{\hat{M}}_{ij}\cdot q_{j} + \boldsymbol{\hat{M}}_{i(d+1)} \right) + \sum_{j=(d+2)}^{\eta} \boldsymbol{\hat{M}}_{ij}\cdot r^{dec}_{j} + \alpha_i.\beta_1
\label{q_tilda}
\end{equation}




\begin{figure*}[!t]
\normalsize
\setcounter{equation}{6}

\hrulefill

\begin{align*}
q'_i &= \beta_2( N\cdot \boldsymbol{\hat{M}}_{ij} +\boldsymbol{\hat{M}}_{i(d+1)})  - (rand.N)\cdot \boldsymbol{\hat{M}}_{it}  + norm\cdot \boldsymbol{\hat{M}}_{ik} + \alpha_i.\beta_1 + small\ number \ \ \\
& \ \ \ \ \ \ \ \ \ \ [Expanding \ r^{dec}_{j},using \ dummy\  indexes]\\
\end{align*}
\vspace{-3em}
\begin{align*}
&= N \left(\beta_2 \cdot \boldsymbol{\hat{M}}_{ij} - rand.\boldsymbol{\hat{M}}_{it}\right) + \beta_2\cdot \boldsymbol{\hat{M}}_{i(d+1)} + norm\cdot \boldsymbol{\hat{M}}_{ik} + \alpha_i.\beta_1 + small\ number
\label{ERidentity}
\end{align*}
\ \ \ \ \ \ \ \ \ Dividing the above term by N, followed by subtraction of remainder modulo N, evaluates to 

\begin{equation}
\frac{q'_i-mod(q'_i,N)}{N} = \beta_2\cdot\boldsymbol{\hat{M}}_{ij} - rand.\boldsymbol{\hat{M}}_{it} + \left(\frac{norm-mod(norm,N)}{N}\right) \boldsymbol{\hat{M}}_{ik} \label{Final_Expand_R_identity}
\end{equation}

\hrulefill

\begin{align}
(\boldsymbol{p}'_i-\boldsymbol{p}'_j)\cdot\boldsymbol{q}' &= \beta_2[(p_{j1}-p_{i1})2q_1 + \cdots + (p_{jd}-p_{id})2q_d + (\Vert \boldsymbol{p}_i \Vert^{2} - \Vert \boldsymbol{p}_j \Vert^{2})] + \beta_1(\alpha_1-\alpha_2)
\label{eq:known_pln_text_attack_eq}
\end{align}

\hrulefill
\label{proof_lemma}
\end{figure*}


In Equation \ref{q_tilda}, QU knows $q_i'$ but does not have any additional knowledge about the RHS. In an attempt to frame query points such that maximum information about the key can be extracted form it, queries of the following structures can be asked:

$\textbf{Case 1:}$ QU asks DO for an encryption of $\boldsymbol{0}^{d}$ and obtains $\boldsymbol{q}'_i$ as:
\begin{equation}
    q'_i= \beta_2\cdot \boldsymbol{\hat{M}}_{i(d+1)} + \sum_{j=(d+2)}^{(d+2+c+\epsilon)} \boldsymbol{\hat{M}}_{ij}\cdot r^{dec}_{j} + \alpha_i.\beta_1
    \label{null_query}
\end{equation}

This is pseudo random for an attacker. Value of $r^{dec}_{j}$, $\beta_2$ and $\alpha_i.\beta_1$ are generated randomly for each query encryption and generated afresh even if the same query point is asked for encryption for the second time. Without the knowledge of $r^{dec}_{j}$, $\beta_2$ and $\alpha_i.\beta_1$ the QU can't frame any linear equations to recover useful information about the secret key $\boldsymbol{\hat{M}}$. Thus queries of form $\boldsymbol{0}^{d}$ reveals no information about the secret key $\boldsymbol{\hat{M}}$.

$\textbf{Case 2:}$ QU asks DO for encryption of $\boldsymbol{e}_j$ and obtains $\boldsymbol{q}'_i$ as
\begin{equation}
q'_i= \beta_2 \left(\boldsymbol{\hat{M}}_{ij} + \boldsymbol{\hat{M}}_{i(d+1)} \right) + \sum_{j=(d+2)}^{(d+2+c+\epsilon)} \boldsymbol{\hat{M}}_{ij}\cdot r^{dec}_{j} + \alpha_i.\beta_1
\label{identity}
\end{equation}

This is similar to Equation \ref{null_query}. As mentioned earlier use of random ephemeral value $r^{dec}_{j}$, $\beta_2$ and $\alpha_i.\beta_1$ makes $q'_i$ pseudo random. Thus without the knowledge of $r^{dec}_{j}$, $\beta_2$ and $\alpha_i.\beta_1$ recover of any columns of $\boldsymbol{\hat{M}}$ from equation \ref{identity} is impossible.

$\textbf{Case 3:}$ QU asks DO for encryption of $(N\cdot\boldsymbol{e}_j)$ for some large integer $N$ as mentioned in Section \ref{attack_mod} and receives $\boldsymbol{q}'_i$ as
\begin{equation}
q'_i= \beta_2 \cdot\left(N\cdot\boldsymbol{\hat{M}}_{ij} + \boldsymbol{\hat{M}}_{i(d+1)}\right) + \sum_{j=(d+2)}^{(d+2+c+\epsilon)} r^{dec}_{j}\cdot\boldsymbol{\hat{M}}_{ij} + \alpha_i.\beta_1
\label{Nidentity}
\end{equation}

Expanding equation \ref{Nidentity} using dummy indexes gives equation \ref{Final_Expand_R_identity}. In equation \ref{Final_Expand_R_identity}, value of 
$rand$, $\beta_2$, and $\frac{norm-mod(norm,N)}{N}$ changes each time even if same query point is asked for encryption second time, which makes $\frac{q'_i-mod(q'_i,N)}{N}$ completely pseudo-random without the knowledge of $rand$, $\beta_2$, and $\frac{norm-mod(norm,N)}{N}$.  Thus, obtaining any information about the secret matrix $\boldsymbol{\hat{M}}$ from Equation \ref{Nidentity} is infeasible.
\subsubsection{Query privacy against CSP having access to polynomial number of plaintext ciphertext pairs} 


Considering a curious CSP having access to polynomial number of plaintext ciphertext pairs. Using these information the CSP can frame equations as discussed in section III of \cite{LinWeipeng}. However, in the proposed encryption scheme it evaluates to something as that of equation \ref{eq:known_pln_text_attack_eq}. 
In equation \ref{eq:known_pln_text_attack_eq}, without the knowledge of $\beta_1(\alpha_1-\alpha_2)$ and $\beta_2$ the CSP can't frame linear equations to recover the plaintext query $\boldsymbol{q}$ uniquely. Thus, the proposed encryption scheme can preserve \textit{query privacy} even against a curious CSP having access to polynomial number of plaintext ciphertext pairs.  

\section{Performance Evaluation\label{results}}
In this section, we will analyze cost of our proposed encryption scheme and compare its performance with that of the earlier encryption schemes
for varying dimensions and varying number of samples considering synthetic database.

\subsection{Cost Analysis:}
The computational complexity of different steps of the proposed encryption scheme is presented in table \ref{Comp_Complexity}. 

In the key generation step, DO generates different secret points and matrices as discussed in the KeyGen step of section \ref{Proposed_scheme} with an overall computational complexity of KeyGen step evaluating to $O(\eta^2)$-- corresponding to the generation of secret matrix $\boldsymbol{M}$. For the database encryption, DO multiplies $m$ data points-- each enhanced to $\eta$-dimensional point for security then multiplied to secret matrix $\hat{\textbf{M}}^{-1}$. Thus total computational complexity evaluating to $O(m\eta^2)$. 

Query encryption is performed in three steps, with QU performing the first step. In this step, the secret key and public key of the Paillier crypto-system are generated, and the $d$-dimensional query point is encrypted. In Paillier-cryptosystem, key generation, one encryption and one decryption costs $O(1)$, $O(\log n)$ and $O(1)$ computational time. Thus computational complexity of this step evaluates to $O(d\log n)$. DO performs the second step of query encryption. In this step, $(\eta-d)$--variables are encrypted followed by multiplication with secret matrix $\hat{\textbf{M}}$. Thus total computational overhead evaluates to $O(\eta^2+(\eta-d)\log n)$. QU performs the third and final step of query encryption. He removes the Paillier encryption layer of the doubly encrypted query point. Thus, the computational complexity of this step evaluates to $O(\eta)$--with the overall computational cost of QU for query encryption evaluating to $O(d\log n)$.

The computational cost of $k$-NN evaluation using the proposed encryption evaluates to $O(m\eta+m\log k)$ as this requires multiplication of $m$ data points with $\eta$-dimension with a query point of the same dimension and comparing the obtained distance to figure out the $k$-NN.  

Considering the proposed encryption scheme's communication cost, DO performs the initial key generation and encryption. Thus zero communication cost involves in these two steps. Considering each tuple of Paillier encrypted query points as $\log n$ bits, the total communication cost between DO and QU sums up to $O(\eta \log n)$ bits. The encrypted query point is then sent to CSP for $k$-NN computation with a communication cost of $O(\eta c)$ bits, where $c$ is the bit length of the encrypted query tuple. Finally, CSP returns computed $k$-NN result with communication cost equivalent to $O(kc)$ bits. Thus totaling communication cost comes out to be $O(\eta+k)c$ bits.

\begin{table*}[h]
    \centering
    \begin{tabular}{|p{2.5cm}|p{2.6cm}|p{1.2cm}|p{1.5cm}|p{2.6cm}|}
         \hline
         Process & DO & QU & CSP & Total \\
         \hline
         Key Generation       & $O(\eta^2)$ & O(1) & - & $O(\eta^2)$   \\
         Database Encryption  & $O(m\eta^2)$ & - & - & $O(m\eta^2)$ \\
         Query Encryption     & $O(\eta^2+(\eta-d)\log n)$ & $O(d\log n)$ & - & $O(\eta^2+\eta\log n)$ \\
         $k$-NN Computation   & - & - & $O(m\eta\log k)$ & $O(m\eta\log k)$   \\
         \hline
    \end{tabular}
    \vspace{1em}
    \caption{Computation complexity of the proposed encryption scheme}
    \label{Comp_Complexity}
\end{table*}

\subsection{Experimental Setup:}
We have implemented the encryption scheme proposed in section \ref{Proposed_scheme}, Zhu et al. \cite{ZhuHuang}, and Sing et al. \cite{SinghKaul} in python, considering a scaling factor of $10^{3}$ to allow real numbers. The security parameters have been fixed to $c=5$ and $\epsilon=5$ for Zhu et al. and Sing et al. however for the proposed encryption scheme  $\epsilon=10$ have been considered.
Paillier encryption scheme with a key size of $1024$ bits is used to maintain query privacy between QU and DO. All experiments are performed on an Intel Core $i7$ $3.40$ GHz CPU  with $8$GB RAM running Ubuntu $18.04$.

\subsection{Experiments performed:\label{per_proposed_scheme}}

\begin{figure}[h!]
  \centering
  \includegraphics[width=9cm, height=5cm]{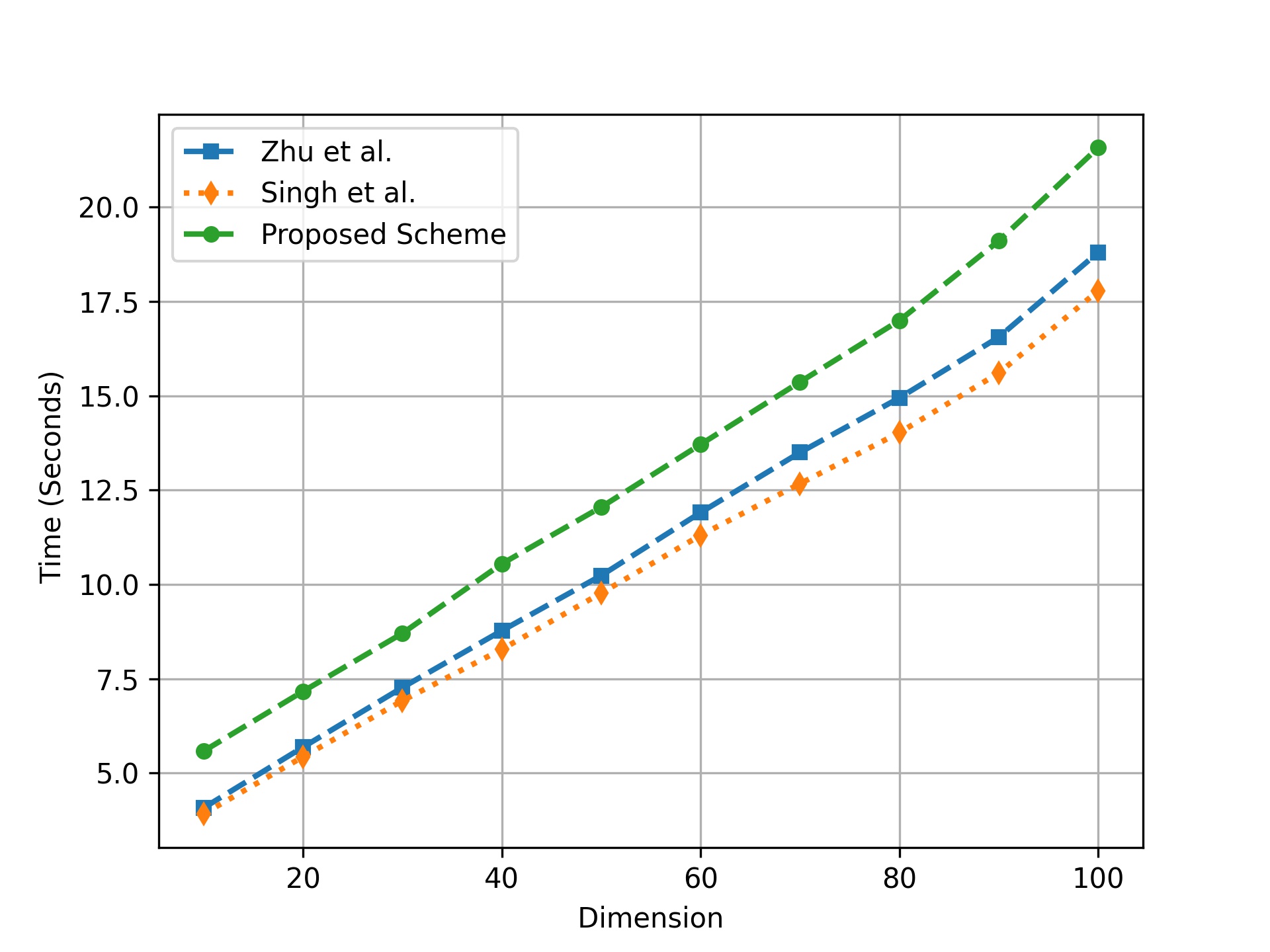}
  \caption{Encryption time of the proposed scheme with that of earlier encryption schemes varying dimension (Data samples=100,000)}
  \label{Encryption_time_dimension}
\end{figure}

Encryption time-varying dimension:
To compare the encryption time of the proposed encryption scheme with varying dimensions, we kept the number of data samples constant at $100,000$ and varied its dimension in the step of $10$. The behavior of different encryption schemes is plotted as shown in figure \ref{Encryption_time_dimension}. All encryption scheme follow a linear trend. 
However, the proposed encryption scheme's computational time is marginally higher than that of the earlier encryption schemes, which is justified as it uses more auxiliary variables for security, i.e., high dimensional $\tau$ point.

\begin{figure}[h!]
    \centering
  \includegraphics[width=9cm, height=5cm]{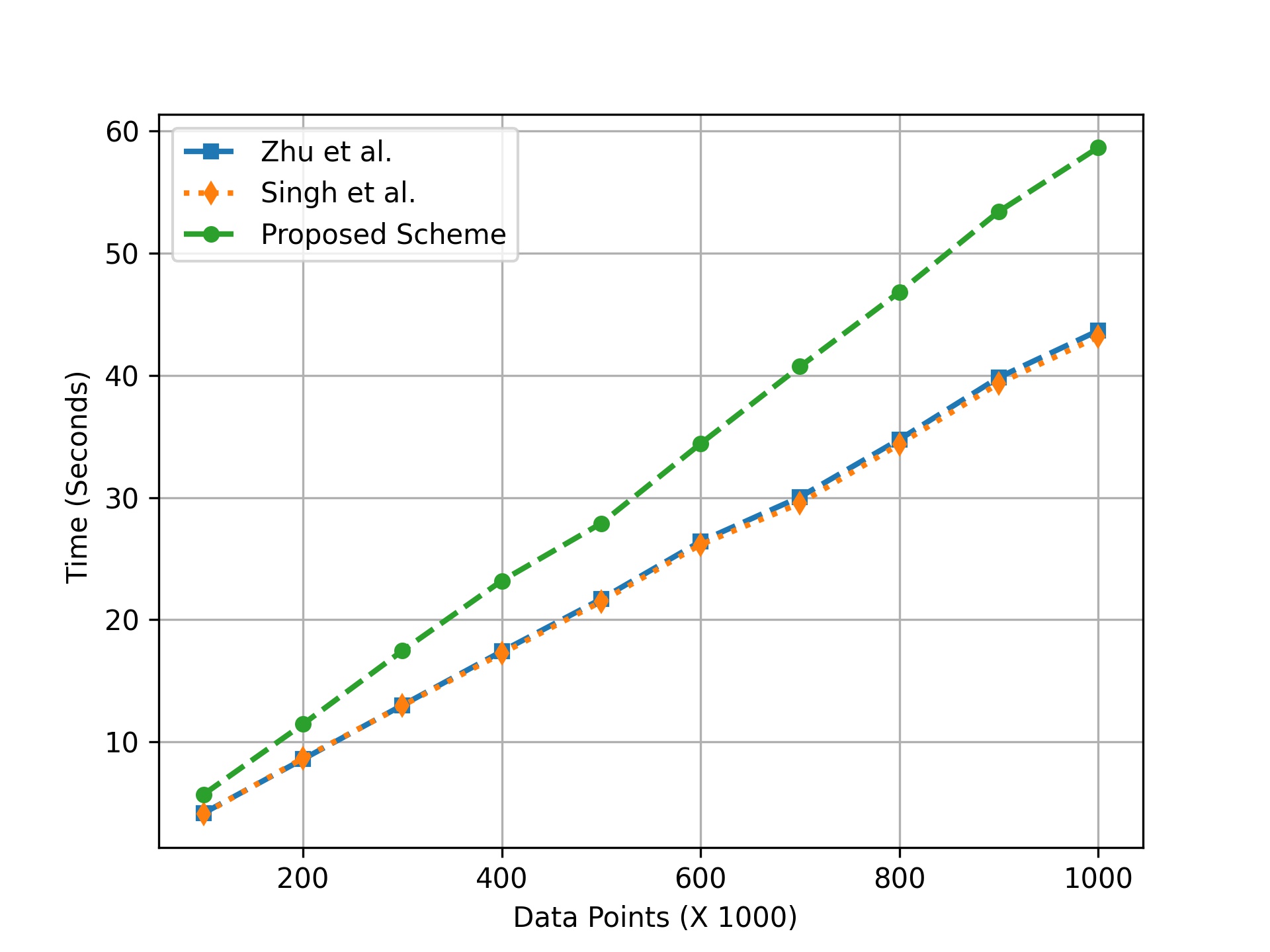}
  \caption{Encryption time of the proposed scheme with that of earlier encryption schemes varying data point dimension (d=$10$)}
  \label{Encryption_time_samples}
\end{figure}

Encryption time varying number of samples: With a varying number of samples keeping data dimension set to $10$ -- a similar trend is observed. Figure \ref{Encryption_time_samples} shows the encryption time with respect to number of samples. An increase in the number of samples results in a linear increase in computation time, with the proposed encryption scheme taking a little bit more time than that of the existing encryption schemes. This is justified as the proposed encryption scheme uses more auxiliary variables than that of the earlier encryption schemes.

\begin{figure}[h!]
    \centering
  \includegraphics[width=9cm, height=5cm]{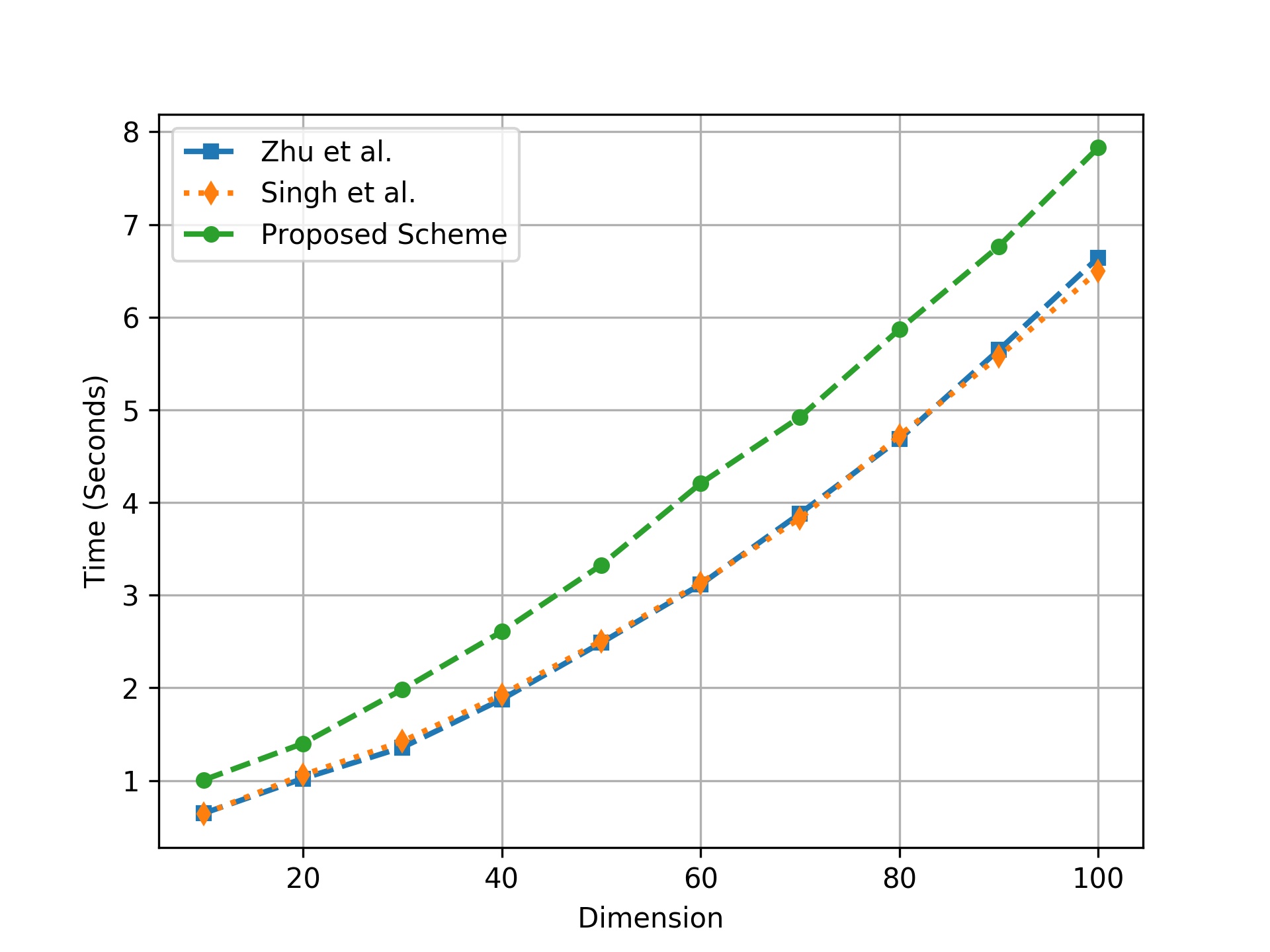}
  \caption{Query encryption time of the proposed scheme with that of earlier encryption schemes varying dimension}
  \label{query_time_dimension}
\end{figure}

Query encryption time varying dimension: Figure \ref{query_time_dimension} shows the comparison of query encryption time of the proposed encryption scheme with that of the earlier encryption schemes. Increase in query dimensionality results in linear increase in query encryption time with a slightly higher encryption time for the proposed encryption scheme.
The increase in computational overhead of the proposed encryption scheme
is, however justified as the proposed encryption scheme requires constant multiplication in the encrypted domain to provide  
key confidentiality against the QU -- property existing schemes can not hold. 

\begin{figure}[h!]
    \centering
  \includegraphics[width=9cm, height=5cm]{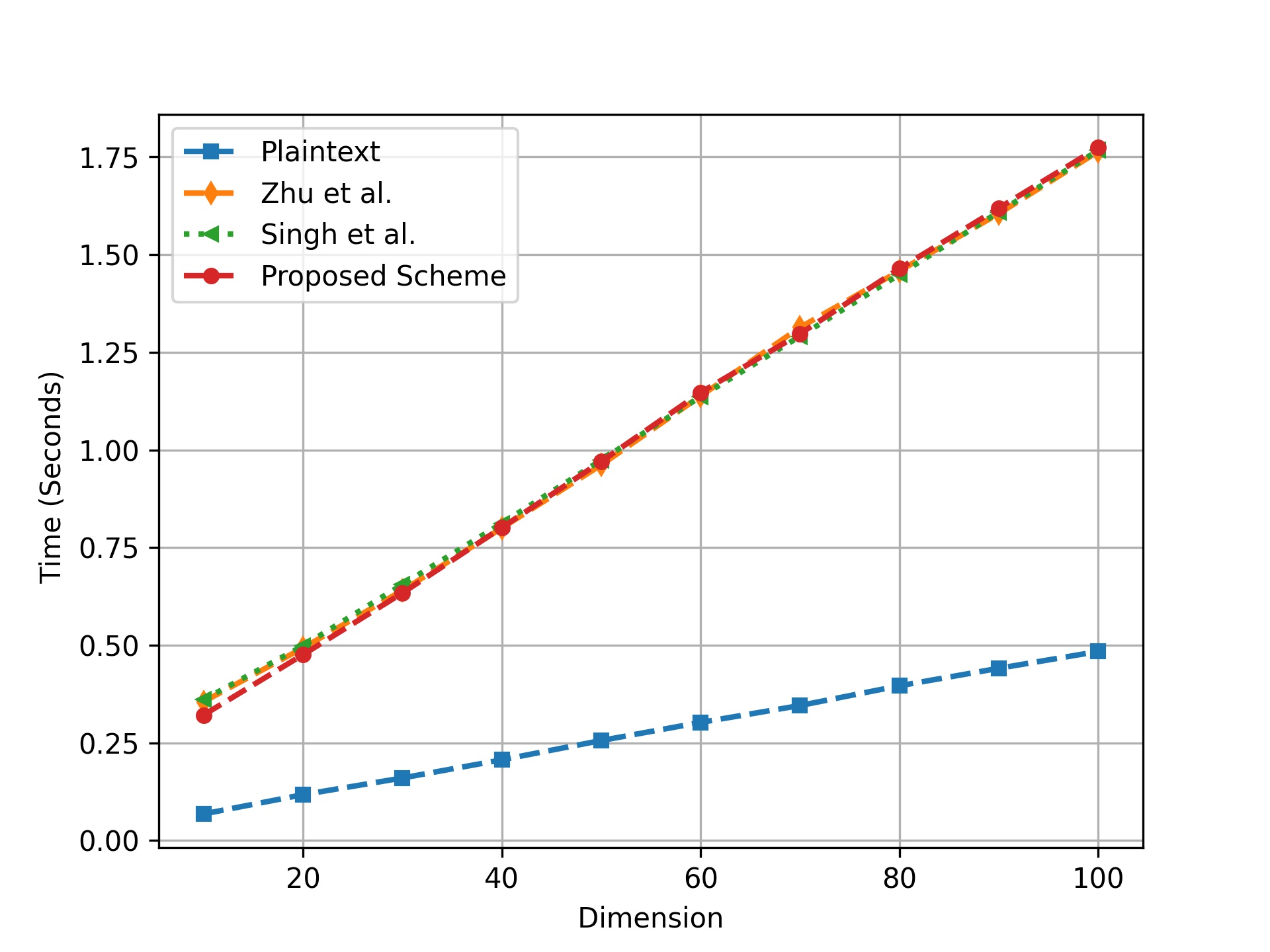}
  \caption{Computation time of $k$-nearest neighbor varying dimension (data samples= $100,000$)}
  \label{kNN_query_dimension}
\end{figure}

\begin{figure}[h!]
    \centering
  \includegraphics[width=9cm, height=5cm]{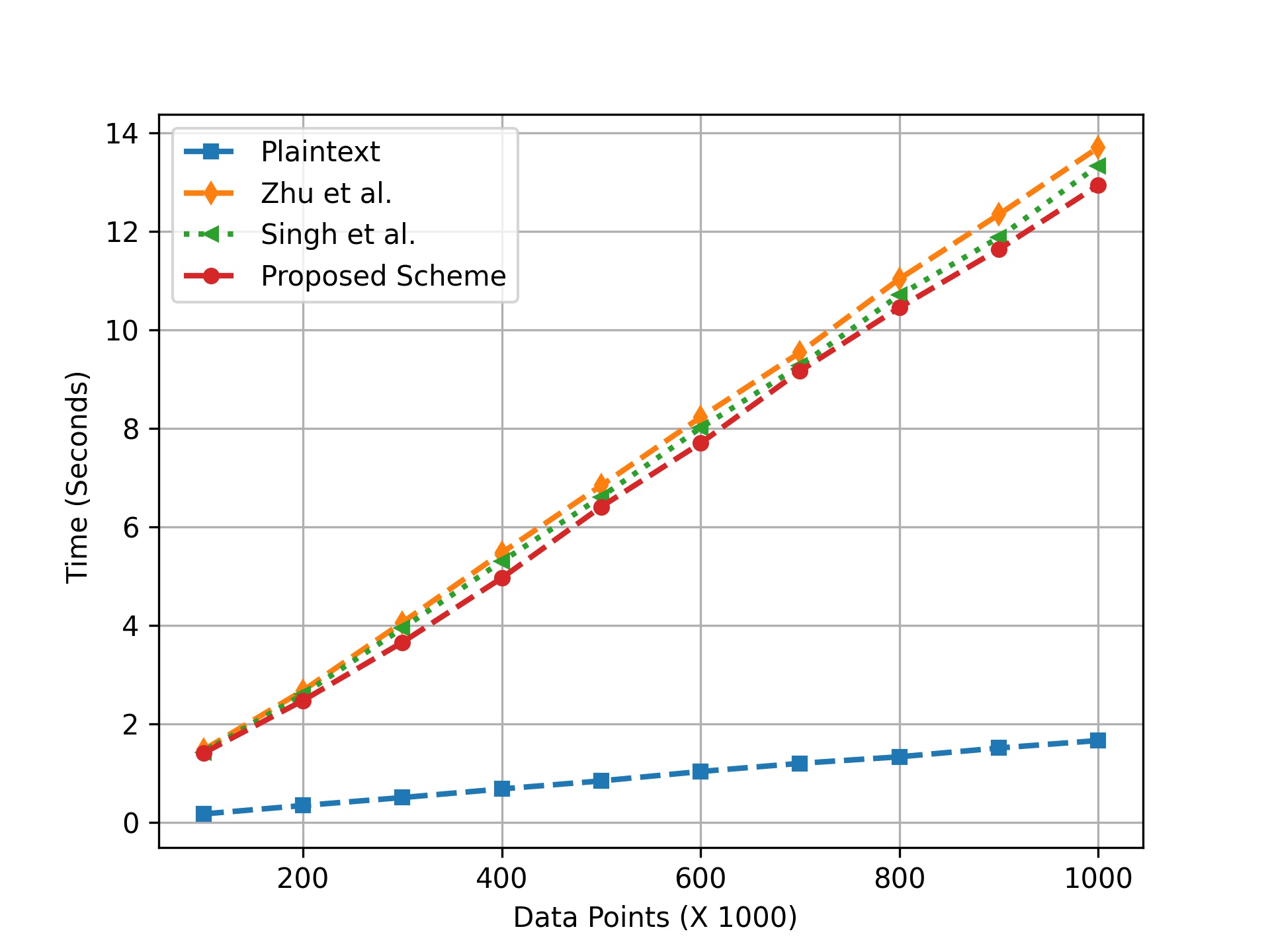}
  \caption{Computation time of $k$-nearest neighbor varying number of data samples keeping constant dimension $d=10$}
  \label{kNN_query_sample}
\end{figure}

$k$-NN computation varying dimension and number of samples: Figure \ref{kNN_query_dimension} and \ref{kNN_query_sample} shows the secure $k$-NN computation in the CSP on encrypted environment. These graphs are sufficient to conclude that $k$-NN computation time varying the number of dimensions and number of samples have the same effect in the proposed encryption scheme as that of the earlier encryption schemes. 

From figure \ref{kNN_query_dimension} and \ref{kNN_query_sample} it is clear that proposed encryption scheme's $k$-NN computation time remains almost similar if not marginally lesser than that of its predecessors. The proposed encryption scheme's performance is comparable to the state-of-art encryption schemes with added key confidentiality, data privacy, and query privacy.


\section{Related Work}
\label{Rel_work}
Searchable encryption is a promising technique to replace the trivial method of downloading the encrypted outsourced data and then decrypting it to search and perform the required operation. 
Searchable Encryption schemes can be divided into two categories: Asymmetric Searchable Encryption (ASE) and Symmetric Searchable Encryption (SSE). The pioneering work proposed by Boneh et.al.\cite{Boneh_Dan_Giovanni} is the first public-key encryption scheme that supports a single keyword search.
However, the ASE schemes are less efficient than SSE schemes due to its complex encryption procedures. 



The concept of SSE is first proposed by Song et al. \cite{Song_Dawn_Xiaoding}.
Later, an SSE-based ranked search over encrypted data is proposed by Wang et al. \cite{Wang_Cong_Ning_Cao}. 
Secure $k$-NN, a key module in many data mining tasks, is an important database analysis operation. 
Wong first proposed the asymmetric scalar product-preserving encryption (ASPE) \cite{WongCheung} to conduct the $k$-NN search over an encrypted database which is viewed as the original secure $k$-NN scheme. Since then, ASPE, with an excellent encryption property of high security and low cost, has been widely studied in many applications \cite{Xu_Zhiyong,Yu_Jiadi_Peng_Lu,Li_Hongwei_Yi_Yang}. ASPE uses an invertible matrix as an encryption/decryption key for data and query points. The basic ASPE scheme is proved to be secure against the level-2 attack. Hence, the author used random asymmetric splitting with additional artificial dimensions to propose an enhanced scheme to resist the level-3
attack. However, recently Weipeng et al. \cite{LinWeipeng} showed the possible level-3 attack in the original Wang's ASPE scheme.



Over time several methods to perform secure $k$-NN computation have been proposed \cite{Xu_Huiqi_Shumin_Guo,Yao_Bin_Feifei_Li,Yiu_Man_Lung_Ira_Assent} are a few of them. Yiu et al. \cite{Yiu_Man_Lung_Ira_Assent} presented three transformation techniques providing some trade-offs among data privacy, query cost, and accuracy. Xu et al. \cite{Xu_Huiqi_Shumin_Guo} proposed a technique to compute a secure query based on the random space perturbation (RASP) method. Yao et al. \cite{Yao_Bin_Feifei_Li} designed a secure $k$-NN query approach based on the partition-based secure Voronoi diagram (SVD). Latter, Choi et al. \cite{Choi_Sunoh_Gabriel_Ghinita} proposed two schemes (VD-kNN and TkNN), which are based on mutable order-preserving encoding (mOPE). The similarity of the works presented above is that each of them has assumed QUs are trusted and DO share his secret key with each of them, which may not be practical and brings several security issues as mentioned in \cite{ZhuTakagi}.

To encrypt query point without sharing DO’s secret key, Zhu et al. \cite{ZhuXu,ZhuTakagi} proposed an interactive protocol for query encryption between DO and QU, such that a QU can only derive partial information from DO’s secret key. Zhu et al. \cite{ZhuHuang} proposed a better encryption scheme applying the Paillier cryptosystem to protect the DO’s secret key for the query encryption. The use of the Paillier cryptosystem helps to achieve DO’s key confidentiality against QUs; however, it requires DO’s online participation's during the query encryption phase. 


Considering the realistic assumption of potential collusion between CSP and QUs, Yuan et al. \cite{Yuan_Jiawei} proposed an efficient encryption scheme which they claimed to resist the level-3 attack even if CSP colludes with some of the QUs. Unfortunately, as pointed out by \cite{Zhu_Youwen_Rong}, a QU-CSP collusion attack can fast break the database security of \cite{Yuan_Jiawei}. Additionally, the scheme in \cite{Yuan_Jiawei} do not preserve query privacy as plain query points are submitted to DO. 


Some research also has been done in the direction of hiding the database access patterns, Elmehdwi et.al \cite{Elmehdwi} introduced the use of Paillier cryptosystem and utilized two non-colluding semi-honest clouds to perform the encrypted $k$-NN query. Based on \cite{Elmehdwi}, Samanthula et.al.\cite{Samanthula} proposed a privacy-preserving $k$-NN classification scheme over the encrypted cloud database capable of hiding the database access patterns. However, both of these schemes assume that DO shares its secret key with one of the clouds, which may not be acceptable in practice and may pose a severe threat to database security and query privacy. Rong et al. \cite{Rong_Hong} proposed a set of secure building blocks and outsourced collaborative $k$-NN protocol in multiple cloud environments. The scheme is based on an ElGamal cryptosystem designed without bit-decomposition, which makes it more efficient than that of the scheme in \cite{Samanthula}.

In the proposed encryption scheme, we are not particularly interested in hiding the database access patterns; thus, we have not included the same in the current context. However, we believe that the same can be achieved using some intelligent techniques.

\section{Conclusion\label{conclusion}}
Computing $k$-NN of a given point is a widely encountered problem in data mining and machine learning. It is of recent interest to compute $k$-NN in a cloud computing scenario where the database is encrypted to preserve data privacy. Out of the several proposed encryption schemes, we have targeted the scheme proposed by Zhu \cite{ZhuHuang} and demonstrated an attack to break the \textit{key confidentiality} claim of the encryption scheme. By issuing a limited number of specially crafted queries, we have shown an attack on recovering a part of the secret key used to encrypt the database. Further, considering a \textit{Level 2 attack scenario} enabled us to retrieve the entire encrypted database. We then presented an enhanced scheme that guarantees \textit{key confidentility}, \textit{data privacy} and \textit{query privacy} with small computational overhead.

\vspace{2em}

\begin{thebibliography}{1}
\bibitem{ZhuTakagi}
Zhu, Youwen, Rui Xu, and Tsuyoshi Takagi. "Secure k-NN computation on encrypted cloud data without sharing key with query users." In Proceedings of the 2013 international workshop on Security in cloud computing, pp. 55-60. 2013.

\bibitem{ZhuXu}
Zhu, Youwen, Rui Xu, and Tsuyoshi Takagi. "Secure k-NN query on encrypted cloud database without key-sharing." International Journal of Electronic Security and Digital Forensics 5, no. 3-4 (2013): 201-217.

\bibitem{ZhuHuang}
Zhu, Youwen, Zhiqiu Huang, and Tsuyoshi Takagi. "Secure and controllable k-NN query over encrypted cloud data with key confidentiality." Journal of Parallel and Distributed Computing 89 (2016): 1-12.

\bibitem{SinghKaul}
Singh, Gagandeep, Akshar Kaul, and Sameep Mehta. "Secure k-NN as a Service Over Encrypted Data in Multi-User Setting." In 2018 IEEE 11th International Conference on Cloud Computing (CLOUD), pp. 154-161. IEEE, 2018.

\bibitem{Elmehdwi}
Elmehdwi, Yousef, Bharath K. Samanthula, and Wei Jiang. "Secure k-nearest neighbor query over encrypted data in outsourced environments." In 2014 IEEE 30th International Conference on Data Engineering, pp. 664-675. IEEE, 2014.

\bibitem{Samanthula}
Samanthula, Bharath K., Yousef Elmehdwi, and Wei Jiang. "K-nearest neighbor classification over semantically secure encrypted relational data." IEEE transactions on Knowledge and data engineering 27, no. 5 (2014): 1261-1273.

\bibitem{Rong_Hong}
Rong, Hong, Hui-Mei Wang, Jian Liu, and Ming Xian. "Privacy-preserving k-nearest neighbor computation in multiple cloud environments." IEEE Access 4 (2016): 9589-9603.

\bibitem{Yuan_Jiawei}
Yuan, Jiawei, and Shucheng Yu. "Efficient privacy-preserving biometric identification in cloud computing." In 2013 Proceedings IEEE INFOCOM, pp. 2652-2660. IEEE, 2013.

\bibitem{Zhu_Youwen_Rong}
Zhu, Youwen, Tsuyoshi Takagi, and Rong Hu. "Security analysis of collusion-resistant nearest neighbor query scheme on encrypted cloud data." IEICE TRANSACTIONS on Information and Systems 97, no. 2 (2014): 326-330.

\bibitem{Boneh_Dan_Giovanni}
Boneh, Dan, Giovanni Di Crescenzo, Rafail Ostrovsky, and Giuseppe Persiano. "Public key encryption with keyword search." In International conference on the theory and applications of cryptographic techniques, pp. 506-522. Springer, Berlin, Heidelberg, 2004.

\bibitem{Song_Dawn_Xiaoding}
Song, Dawn Xiaoding, David Wagner, and Adrian Perrig. "Practical techniques for searches on encrypted data." In Proceeding 2000 IEEE Symposium on Security and Privacy. S\&P 2000, pp. 44-55. IEEE, 2000.

\bibitem{Wang_Cong_Ning_Cao}
Wang, Cong, Ning Cao, Kui Ren, and Wenjing Lou. "Enabling secure and efficient ranked keyword search over outsourced cloud data." IEEE Transactions on parallel and distributed systems 23, no. 8 (2011): 1467-1479.

\bibitem{Xu_Zhiyong}
Xu, Zhiyong, Wansheng Kang, Ruixuan Li, KinChoong Yow, and Cheng-Zhong Xu. "Efficient multi-keyword ranked query on encrypted data in the cloud." In 2012 IEEE 18th International Conference on Parallel and Distributed Systems, pp. 244-251. IEEE, 2012.

\bibitem{Yu_Jiadi_Peng_Lu}
Yu, Jiadi, Peng Lu, Yanmin Zhu, Guangtao Xue, and Minglu Li. "Toward secure multikeyword top-k retrieval over encrypted cloud data." IEEE transactions on dependable and secure computing 10, no. 4 (2013): 239-250.

\bibitem{Li_Hongwei_Yi_Yang}
Li, Hongwei, Yi Yang, Tom H. Luan, Xiaohui Liang, Liang Zhou, and Xuemin Sherman Shen. "Enabling fine-grained multi-keyword search supporting classified sub-dictionaries over encrypted cloud data." IEEE Transactions on Dependable and Secure Computing 13, no. 3 (2015): 312-325.



\bibitem{WongCheung}
Wong, Wai Kit, David Wai-lok Cheung, Ben Kao, and Nikos Mamoulis. "Secure kNN computation on encrypted databases." In Proceedings of the 2009 ACM SIGMOD International Conference on Management of data, pp. 139-152. ACM, 2009.

\bibitem{Yiu_Man_Lung_Ira_Assent}
Yiu, Man Lung, Ira Assent, Christian S. Jensen, and Panos Kalnis. "Outsourced similarity search on metric data assets." IEEE Transactions on knowledge and data engineering 24, no. 2 (2010): 338-352.

\bibitem{Xu_Huiqi_Shumin_Guo}
Xu, Huiqi, Shumin Guo, and Keke Chen. "Building confidential and efficient query services in the cloud with RASP data perturbation." IEEE transactions on knowledge and data engineering 26, no. 2 (2012): 322-335.

\bibitem{Yao_Bin_Feifei_Li}
Yao, Bin, Feifei Li, and Xiaokui Xiao. "Secure nearest neighbor revisited." In 2013 IEEE 29th international conference on data engineering (ICDE), pp. 733-744. IEEE, 2013.

\bibitem{Choi_Sunoh_Gabriel_Ghinita}
Choi, Sunoh, Gabriel Ghinita, Hyo-Sang Lim, and Elisa Bertino. "Secure knn query processing in untrusted cloud environments." IEEE Transactions on Knowledge and Data Engineering 26, no. 11 (2014): 2818-2831.






\bibitem{Paillier}
Paillier, Pascal. "Public-key cryptosystems based on composite degree residuosity classes." In International Conference on the Theory and Applications of Cryptographic Techniques, pp. 223-238. Springer, Berlin, Heidelberg, 1999.

\bibitem{ChenLiu} 
Chen, Keke, and Ling Liu. "Privacy preserving data classification with rotation perturbation." In Fifth IEEE International Conference on Data Mining (ICDM'05), pp. 4-pp. IEEE, 2005.

\bibitem{DaemenRijmen}
Daemen, Joan, and Vincent Rijmen. The design of Rijndael: AES-the advanced encryption standard. Springer Science \& Business Media, 2013.








\bibitem{Delfs}
Delfs, Hans, Helmut Knebl, and Helmut Knebl. Introduction to cryptography. Vol. 2. Heidelberg: Springer, 2002.

\bibitem{LinWeipeng}
Lin, Weipeng, Ke Wang, Zhilin Zhang, and Hong Chen. "Revisiting security risks of asymmetric scalar product preserving encryption and its variants." In 2017 IEEE 37th International Conference on Distributed Computing Systems (ICDCS), pp. 1116-1125. IEEE, 2017.


\end{thebibliography}
\end{document}